\documentclass{aa}
\usepackage{graphicx}
\usepackage{txfonts}

\begin{document}

\title{Magnetic fields and gas in the cluster influenced spiral galaxy\\
NGC\,4254 -- I. Radio and X-rays observations\\
}

\author{K. T. Chy\.zy\inst{1},
  M. Ehle\inst{2}, and
  R. Beck\inst{2}} 
\institute{Astronomical Observatory, Jagiellonian
University, ul. Orla 171, 30-244 Krak\'ow, Poland
 \and XMM-Newton Science Operations Centre, ESA, 
 P.O. Box 78, 28691 Villanueva de la Ca\~nada, Madrid, Spain
 \and Max-Planck-Institut f\"ur Radioastronomie, Auf dem H\"ugel 69, 
53121 Bonn, Germany
  }

\date{Received 15/06/2007/ Accepted 06/08/2007}

\titlerunning{Magnetic fields and gas in NGC\,4254}
\authorrunning{K. T. Chy\.zy et al.}

\abstract
{}
{Radio observations can show how cluster galaxies are affected by various 
environmental factors, which perturb their morphology as well as modify properties of 
the interstellar medium (ISM), especially its magnetic field characteristics.
}
{We made high--resolution and high--sensitivity radio--polarimetric VLA 
observations of NGC\,4254 at three frequencies (8.46, 4.86 and 1.43\,GHz). The 
interferometric data were extended with single--dish 
(100--m Effelsberg) observations. Next we performed sensitive 
XMM--Newton observations in X--rays and UV light to investigate the hot gas component and 
its possible interaction with the hot cluster medium. For a complete picture of 
the interplay between various gas phases, we also used optical, \ion{H}{i} and 
infrared (Spitzer) data.}
{The distribution of total radio intensity at 8.46\,GHz and 4.86\,GHz
reveals a global asymmetry with a more diffuse and almost two times 
larger extension to the north than to the south. The radio polarized intensity 
is even more asymmetric, showing a strange bright ridge in the southern 
disk edge, displaced to the downstream side of the local density wave. Magnetic 
arms can be also seen in other disk portions, mostly avoiding 
nearby optical spiral arms. Spatially resolved emission of hot X--ray gas from the 
whole galactic disk, with its soft component closely tracing star--forming regions, is 
detected. Various gas components of thermal origin show strong wavelet 
cross--correlations ($r_w\ge 0.8$), but the polarized intensity anticorrelates 
($r_w=-0.4$) with the thermal and X--ray emission. 
The slope of the local radio nonthermal--infrared 
relation is $<1$, thus smaller than for the radio thermal--infrared one ($\ge1$).
Using the radio thermal emission--based star formation rate (SFR)  
we find higher extinction 
in more H$\alpha$ luminous star--forming regions with a power--law slope of 0.83. 
The galaxy's estimated mean SFR of $0.026\,\mathrm{M}_{\sun}\,
\mathrm{yr}^{-1}\,\mathrm{kpc}^{-2}$ is three times larger than in other spirals of 
similar Hubble type.
}
{ 
NGC\,4254 seems to be a `young' Virgo cluster member, which recently experienced 
a gravitational encounter at the cluster's periphery, which perturbed its spiral 
arms by tidal forces and triggered a burst of star--formation which still 
maintains strong radio and infrared emissions. Tidal forces could 
also sheared the magnetic field in the 
southern disk portion and led to the observed polarized ridge, though, 
magnetic field compression by weak ram pressure forces of the cluster gas cannot be excluded.
The case of NGC\,4254 shows that 
the polarized signal provides additional information on MHD 
processes acting on magnetized plasma during the galaxy's evolution, which cannot 
be obtained from any other ISM component.}

\maketitle

\section{Introduction}
\label{intro}
\begin{figure}

\resizebox{\hsize}{!}{\includegraphics{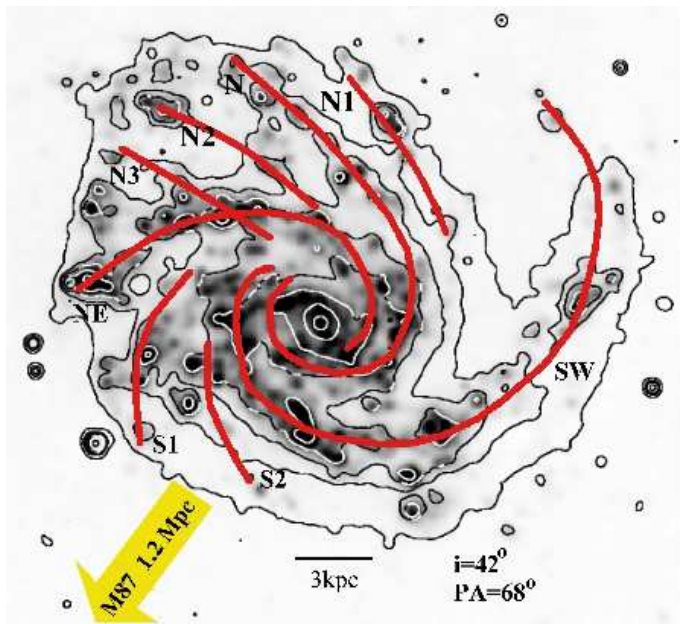}}

\caption{
Contour plot of the NIR IRAC $3.6\,\mu{\rm m}$ emission of the spiral (Sc) galaxy 
NGC\,4254 overlaid on 
H$\alpha$ map at 4\arcsec\ resolution (data from the Spitzer survey of SINGS 
galaxies, Kennicutt et al. \cite{kennicutt03}). The spiral arms 
(SW, N, NE) and arm--like features (N1, N2, N3, and S1, S2) are marked. The linear size 
scale, galaxy inclination, and position angle of the galactic major 
axis are also indicated. 
} 
\label{f:arms}
\end{figure}

Galaxies in clusters are known to be affected by a number of environmental factors, 
which may change the galaxy's star--forming activity, leading to \ion{H}{i} deficiency, 
spatial truncation of H$\alpha$ emitting gas, morphological transformations, and 
strong dynamical galaxy evolution (Boselli \& Gavazzi \cite{boselli06}). The 
ISM of disk galaxies in clusters can be altered by the cluster's ram pressure stripping 
and/or the galaxies' gravitational interactions redistributing the stellar 
as well as the gaseous disk content (Kenney et al. \cite{kenney04}, Vollmer et al. \cite{vollmer01}). 
Investigating such galaxies at various wavelengths enables one to recognize 
particular physical processes at work within the perturbed ISM, and to provide a 
tentative scenario of the galaxy's interaction with the cluster medium (Machacek et 
al. \cite{machacek04}).

High quality observations at radio frequencies are of special interest 
as they enable us to separate the thermal part of radio emission, thus 
providing a precise tracer of star--forming activity within galactic disks, unaffected 
by dust extinction. The polarized radio emission can reveal the regular magnetic 
field and its response to peculiar gas flows, giving an independent insight 
into the disturbing agents in cluster spirals (Chy\.zy et al. \cite{chyzy06}).

The Virgo Cluster is the nearest moderately rich cluster 
with more than a hundred bright spirals and lenticulars. 
It is dynamically young, with infalling galaxies and merging subgroups (Binggeli et 
al. \cite{binggeli93}), while showing, specific for clusters, high velocity 
gravitational encounters and galaxies in the process of morphological transformation. 
Galaxies of disturbed H$\alpha$ disks (Koopman \& Kenney \cite{koopmann04}) 
or \ion{H}{i} gas depletion (Cayatte et al. \cite{cayatte94}) are ideal to study 
in detail the environmental processes affecting the galaxy evolution.

The aim of this paper is to investigate 
the influence of the intracluster environment on the 
properties of radio emission and magnetic fields in the weakly disturbed 
Virgo Cluster spiral galaxy NGC\,4254. The galaxy shows some external influence 
by its unusual optical spiral pattern (Fig.~\ref{f:arms}), dominated by a strange 
three--arm structure (SW, N, NE, Gonz\'ales \& Graham \cite{gonzales}). 
The southern arm is particularly bright and thick with ongoing 
vigorous star formation. The northern and eastern part of the galaxy show a 
number of weaker flocculent features, which makes the galaxy to appear asymmetrical 
at optical images. 

NGC\,4254 is located far away from the Virgo centre -- 3.7$\degr$ 
(1.2\,Mpc) in the NW direction from M\,87 (Fig.~\ref{f:arms}). Its radial velocity of 
2407\,km\,s$^{-1}$ (Phookun et al. \cite{phookun93}) when compared to the mean 
heliocentric velocity of the cluster of $1050\pm35$\,km\,s$^{-1}$ (Binggeli et 
al. \cite{binggeli93}) gives an additional line--of--sight velocity 
component of about 1400\,km\,s$^{-1}$. This value is approximately two times 
larger than the velocity dispersion of the cluster spirals, which can indicate that, 
if the galaxy has an orbital velocity typical for the Virgo Cluster galaxies, 
its actual velocity vector is almost tangential to the line of sight. As compared to 
the other cluster members the galaxy shows a rather small \ion{H}{i} 
deficiency (0.17, Cayatte et al. \cite{cayatte94}), in agreement with the peripheral
location of the galaxy within the cluster. However, the 
\ion{H}{i} distribution of NGC\,4254 is definitely asymmetric (Phookun et 
al. \cite{phookun93}) with a long extension to the north, and  
a one--arm dominated spiral pattern. It was proposed that the perturbed 
\ion{H}{i} distribution could have been caused by tidal forces (Minchin et al. 
\cite{minchin}) or tidal as well as ram pressure forces, due to the galaxy motion 
through the intracluster medium (ICM) (Vollmer et al. \cite{vollmer05}). 

Only low--resolution single dish radio observations of NGC\,4254 at 
10.45\,GHz were available up to now (Soida et al. \cite{soida}). They show some 
deviations in the total radio intensity from a symmetric disk as well as
regions of strong polarized emission.  
To investigate in detail the influence of the cluster medium on radio emission 
and magnetic field we obtained sensitive three--frequency VLA radio 
polarimetric data. The interferometric data were merged with single--dish 
Effelsberg observations to recover the diffuse large--scale emission.  
In order to study the hot medium, we performed XMM--Newton observations 
in X--rays and UV light. For a more complete picture 
we also re--processed the \ion{H}{i} data from the VLA public archive, as well as 
gathered H$\alpha$, optical (including HST) and infrared images
from the Spitzer survey of SINGS galaxies (Kennicutt et al. \cite{kennicutt03}).

Spatial comparisons of radio total, thermal, and synchrotron 
components with the other ISM species are performed by the wavelet cross--correlation 
method (Sect.~\ref{s:wave}). We analyze the current state of star--forming 
processes by establishing the distribution of extinction--free star formation rate
(SFR) {\em within} the galaxy, based on the radio thermal emission data
(Sect.~\ref{s:sfr}). We also investigate the radio--infrared correlation diagram 
constructed for the regions within the galaxy as well as locate the galaxy 
on the global radio--infrared relation for a sample 
of non--interacting field spirals (Sect.~\ref{s:fir}). Finally, we estimate 
the relative importance of various competing internal and external forces at work in  
ISM by comparing the corresponding energy densities in the 
galactic disk and discuss possible interaction scenarios 
for the galaxy (Sect.~\ref{s:scenarios}).
  
In the sequel Paper II (Chy\.zy \cite{paperII}) we will investigate the 
Faraday rotation and various depolarization processes in NGC\,4254 and 
perform an analytical modelling of the impact of ram--pressure and tidal 
forces on the galactic magnetic field.

\section{Radio and X--ray observations and data reduction}
\label{s:obs}

\subsection{VLA and Effelsberg data}
\label{vla}

\begin{table*}
 \caption[]{Parameters of VLA observations and exemplary resolution and 
 rms noise in the final maps of I, Q, U Stokes parameters.}
\label{t:vla}
 \centering
 \begin{tabular}{llll}
 \hline\hline
 & Band 1 & Band 2 & Band 3 \\
 \hline
 Frequency [GHz]         & 8.435 \& 8.485 & 4835 \& 4.885 & 1.385 \& 1.465 \\
 Observing date          &  1997 Nov 21; Dec 12 & 1999 May 10 & 1998 Dec 13 \\
 Configuration           & D & D & C\\
 Net observing time [h]  & 13.6 & 7.2 & 5.5 \\
 HPBW/$\sigma_{\rm I},\,\sigma_{\rm Q,U}$ [$\mu$Jy/b.a.] & $10\arcsec$ / 9, 7 & $15\arcsec$ 
/ 10, 9  & $17\arcsec$ / 15, 8  \\
\hline 
\end{tabular}
\end{table*}

High--resolution and high--sensitivity radio polarimetric observations of 
NGC\,4254 were made at 8.46\,GHz, 4.86\,GHz, and 1.43\,GHz using the VLA of 
NRAO\footnote{National Radio Astronomy Observatory is a facility of National 
Science Foundation operated under cooperative agreement by Associated 
Universities, Inc.}. The observations at 8.46\,GHz and 4.86\,GHz were performed 
in the D--array configuration and at 1.43\,GHz with the C--array. All 
observations were carried out in the continuum mode with two independent channels 
of 50\,MHz width (see Table~\ref{t:vla} for details).

The intensity scale at all frequencies was calibrated by observing 3C286.
The position angle of linearly polarized intensity was calibrated using the same source 
with an assumed position angle of 33$\degr$. At 4.86\,GHz and 8.46\,GHz the 
calibrator 1236+077 was used to determine the telescope phases and
instrumental polarization. In order to check these calibration procedures 3C138 was 
observed once each observing day. 

The data reduction was performed using the AIPS data reduction package. 
The visibility data from each observational day were edited, calibrated, 
and self--calibrated in phase. The data from subsequent days were next combined, 
again self--calibrated in phase, and deconvolved to obtain 
maps in Stokes parameters I, Q and U at all three frequencies. 
Adopting different weighting of UV data, with the robustness parameter 
(Briggs \cite{briggs95}) ranging from $-1$ to $+1$, maps with different resolutions 
and sensitivities were obtained. All maps were then corrected for primary beam 
attenuation. Typical rms noise in I, Q, and U maps as well as beam sizes at all frequencies 
are given in Table~\ref{t:vla}.

In order to increase the sensitivity to extended structures, we performed additional 
observations with the 100--m Effelsberg radio telescope\footnote{The 
100--m telescope at Effelsberg is operated by the Max--Planck--Institut f\"ur 
Radioastronomie (MPIfR) on behalf of the Max--Planck--Gesellschaft.} at
4.85\,GHz using the two--horn system. 16 coverages of $40\arcmin\times30\arcmin$ 
size were obtained for each horn during the observations in May and June of 1999. 
We also observed 3C286 to establish the flux 
density scale. The coverages from both the horns were edited by removing spikes of 
interferences and large--scale undulations in baseline levels. Next they were 
combined using the ``software beam switching'' technique (Morsi \&~Reich 
\cite{morsi86}) followed by restoration of I, Q, and U maps. All the appropriate 
maps were then calibrated and combined using the spatial--frequency weighting 
method and digital filtering process to remove the spatial frequencies 
corresponding to structures smaller than the Effelsberg beam ($153\arcsec$ HPBW). 
The rms noise level in the final map of total intensity is 0.5\,mJy/b.a.
and in Q and U maps 0.07\,mJy/b.a.
In a similar way we also re--reduced and calibrated the Effelsberg observations 
of NGC\,4254 at 10.45\,GHz of $65\arcsec$ resolution obtained by Soida et al. 
(\cite{soida}) to ensure that all the data used were fully compatible.

In the UV plane we merged the I, Q, and U -- VLA data at 8.46\,GHz and 4.86\,GHz 
with single--dish observations at 10.45\,GHz and 4.85\,GHz, respectively. The 
brightness values at 10.45\,GHz were rescaled to 8.46\,GHz assuming a spectral 
index of 0.8 ($\propto \nu ^{-\alpha}$). 
The spectral index may vary locally, but the VLA data cover spatial 
frequencies well down to 1.1\,k$\lambda$ (which corresponds to a baseline of 40\,m 
and angular resolution of 3\arcmin), hence the Effelsberg images add only the very 
diffuse signal on larger scales. Taking the inner one--third of the galaxy 
extent and changing the spectral index e.g. to 0.7 would result in the
total signal increasing by 2\%. For the polarized intensity, which is less extended than 
the total one, the errors of similar kind are expected to be even less. As the 
Faraday rotation measures are small in this galaxy (up to about 
$100\,\mathrm{rad}\,\mathrm{m}^{-2}$, Paper II) the rescaling of the 
Effelsberg map may introduce a polarized angle offset up to 3\degr. None 
of those potential errors could change any conclusion of this paper.

We compared different methods
of merging our interferometric and single--dish data using 
AIPS task IMERG and MIRIAD task IMMERGE, which differ in ways of 
weighting the data in the Fourier space. We chose the second 
method, which better filled the regions of negative VLA response around 
the high--gradient emission in the southern outer parts of the galactic disk and
precisely recovered the total flux out of the single dish data.

The merged I, Q, and U maps were then combined to get the 
distributions of total and linearly polarized intensities, as well as the position 
angle of magnetic vectors (not corrected 
for Faraday rotation $\vec{B}$--vectors, i.e. observed $\vec{E}$--vectors rotated by $90\degr$).

The so far published VLA \ion{H}{i} data of NGC\,4254 by Phookun et al. 
(\cite{phookun93}) with velocity resolution of $10\,{\rm km s^{-1}}$ and spatial 
resolution of $35\arcsec$ are not sufficient for a detailed comparison either with our radio
data (with at least $15\arcsec$ resolution) or with the X--ray and
optical images. Therefore, we re--reduced the original VLA archived \ion{H}{i} data from 
the VLA C and D arrays. We used the standard approach to calibration, cleaning  
and combining the data from both arrays, as well as to continuum subtraction procedures 
using AIPS. The final cleaned \ion{H}{i} cube with uniform weighting had the 
resolution of $15\arcsec$ and was used to derive the velocity integrated 
\ion{H}{i} intensity distribution, velocity field and velocity dispersion.

\subsection{XMM--Newton data}
\label{xmmobs}
\subsubsection{X--ray data}

On the 29th June, 2003, we observed NGC\,4254 with XMM--Newton\footnote{
XMM--Newton is an ESA science mission with instruments and contributions directly 
funded by ESA Member States and NASA (Jansen et al. \cite{jansen01})} 
in the energy band of 0.2--12~keV and about $25\arcmin\times 25\arcmin$ 
field of view (Observation Id: 
0147610101). The EPIC MOS cameras (Turner et al. \cite{turner01}) were used in
"Full Frame" and the pn camera (Str\"uder et al. \cite{struder01}) in 
"Extended Full Frame" mode with thin filters for the scheduled observing time of 
52\,ks. The Science Analysis System (SAS) version 6.0.0 was used to calibrate and
analyse the data. As the observations were affected by high radiation, the
standard procedures mentioned in the SAS User's Guide to define good time intervals
were followed. Applying these selections to our data resulted in a clean observing 
time of 18\,ks for MOS and 14.8\,ks for pn. MOS and pn data were exposure--corrected,
merged, and smoothed with a Gaussian filter to $10\arcsec$ HPBW resolution.

A detailed analysis of the XMM--Newton data including a discussion of all X--ray 
emission components and their spectral properties is beyond the scope of this
paper and will be presented in Ehle et al. (in prep.). 

\subsubsection{Optical Monitor data}
\label{uvobs}

During the X--ray observations, NGC\,4254 was also observed with the 
onboard XMM--Newton Optical Monitor (OM, Mason et al. \cite{mason01}) 
in UV filters UVW1 (2905\,\AA), UVM2 (2310\,\AA) and UVW2 (2070\,\AA)
in "full frame low resolution" mode. The exposure times were 3\,ks, 
6\,ks and 18\,ks, respectively. We used the SAS package omichain 
(Mason et al. \cite{mason01}) for photometric reduction and construction of a composite image in each 
of the UV filter bands.

\begin{figure*} [t]
\begin{minipage}[b]{1\linewidth}
\centering
\includegraphics[width=8.9cm]{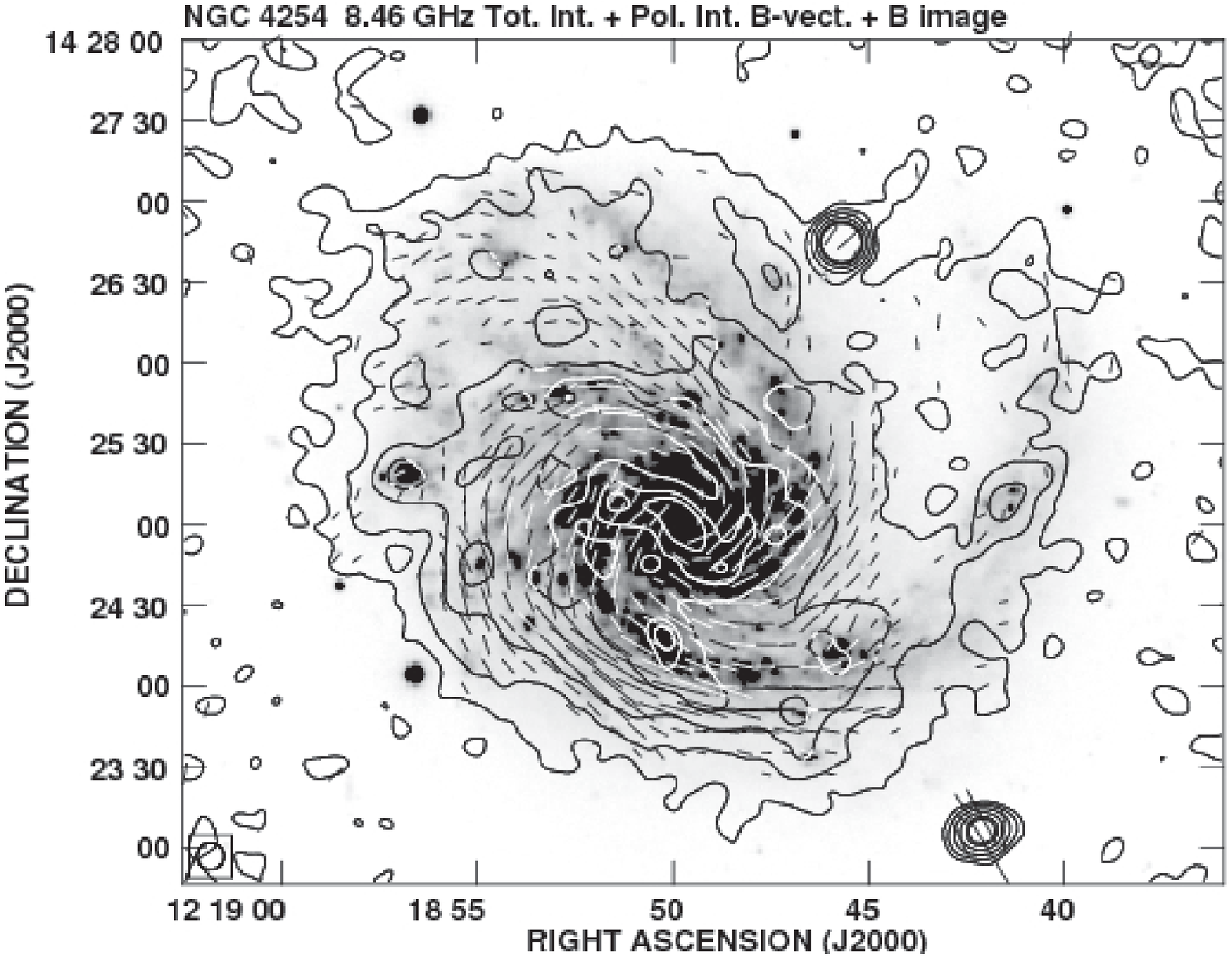}
\includegraphics[width=8.9cm]{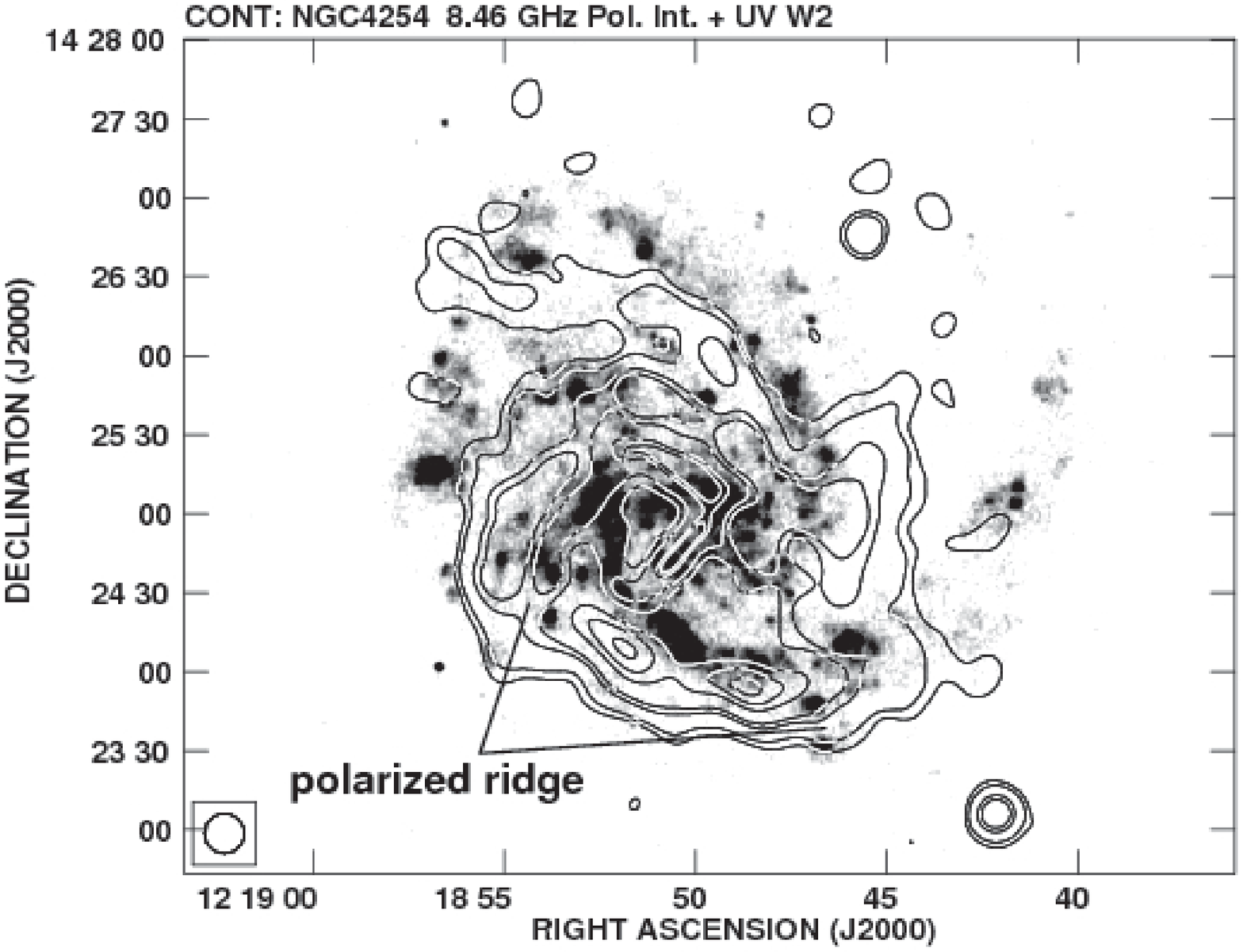}
\end{minipage}\\
\begin{minipage}[b]{1\textwidth}
\centering
\includegraphics[width=8.9cm]{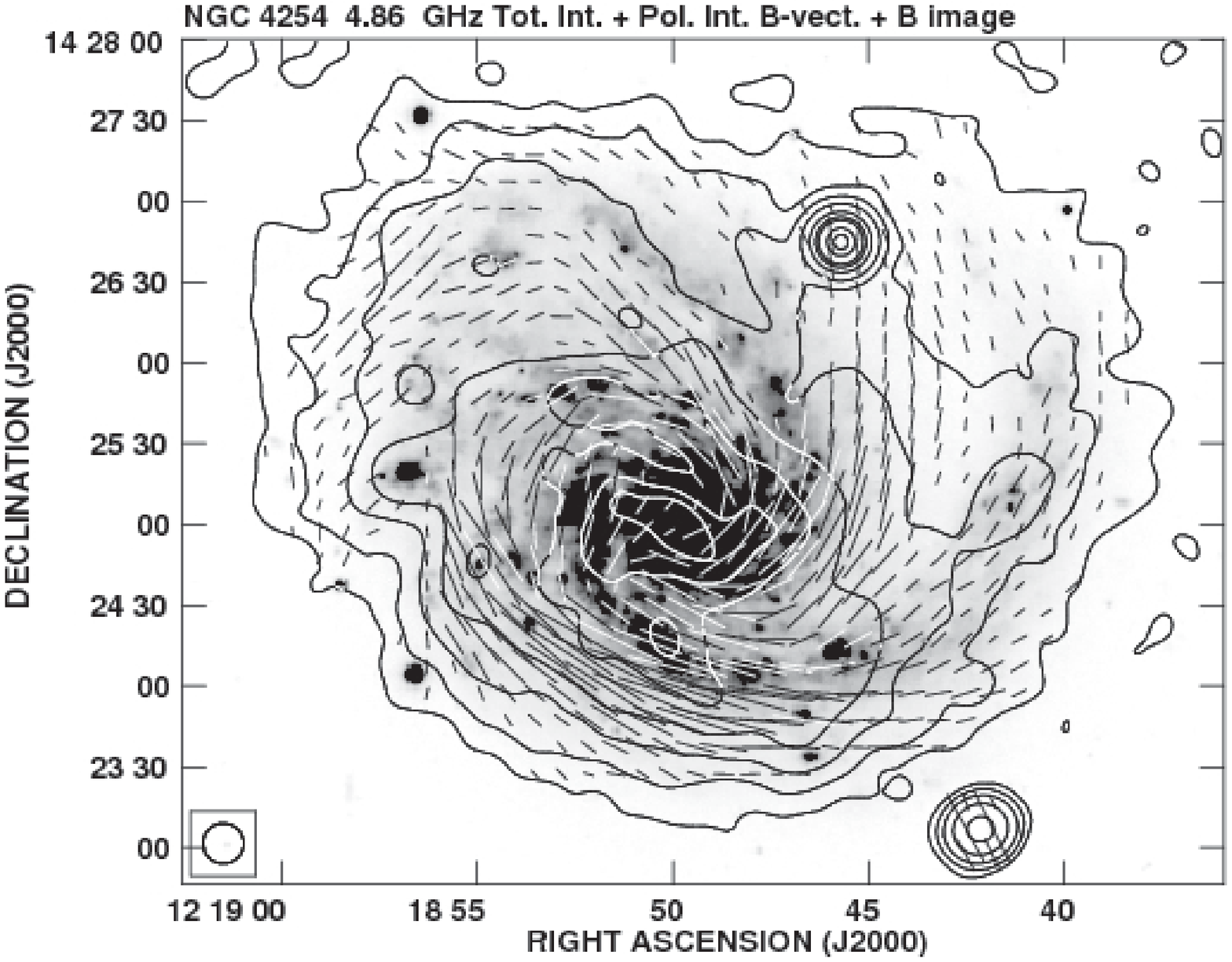}
\includegraphics[width=8.9cm]{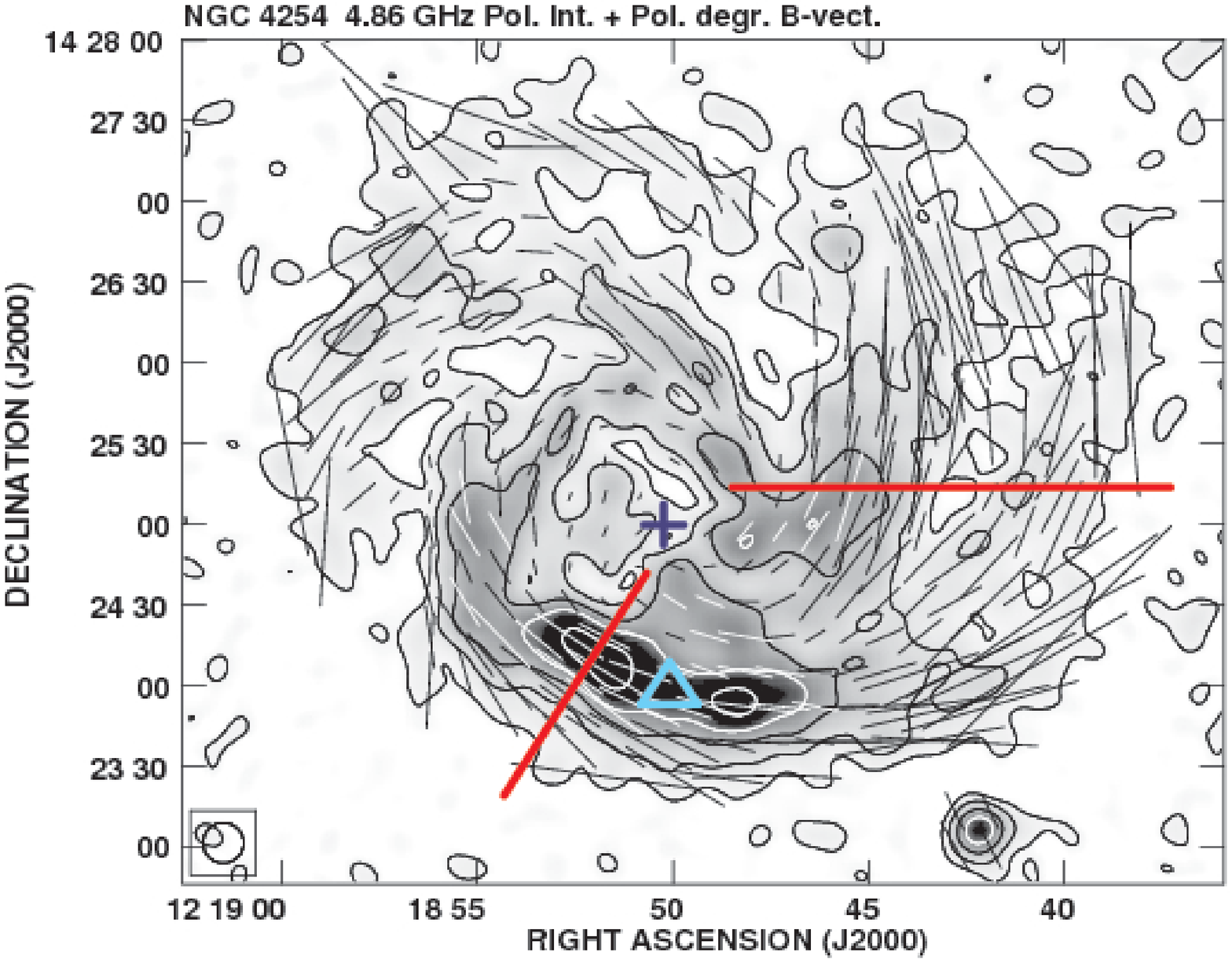}

\caption{Top--left: The contours of total radio intensity of NGC\,4254 at 8.46\,GHz 
with resolution of $10\arcsec$  and $\vec{B}$--vectors (not corrected for Faraday 
rotation) proportional to the polarized intensity overlaid on the 
optical blue image (from Knapen et al. \cite{knapen03}). The contours are at: 
30, 80, 160, 320, 640, 1280 $\mu$Jy/b.a. Vectors of length of $1\arcsec$ 
correspond to polarized intensity of $5\mu\mathrm{Jy/b.a.}$ 
Top--right: The contours of polarized intensity at 8.46\,GHz with a resolution of 
$15\arcsec$ overlaid on the UVW2 image from our XMM satellite observations, 
denoting the ensembles of young stars. Bottom--left: The contours of total 
intensity at 4.86\,GHz with a resolution of $15\arcsec$ with 
$\vec{B}$--vectors proportional to the polarized emission overlaid on the optical 
B--image. The contours are at: 30, 80, 160, 320, 640, 
1280 $\mu$Jy/b.a. Vectors of length of $1\arcsec$ correspond to
a polarized intensity of $8.3\mu\mathrm{Jy/b.a.}$ Bottom--right: The contours and 
greyscale of polarized intensity at 4.86\,GHz 
with a resolution of $15\arcsec$ with $\vec{B}$--vectors of percentage of polarization. 
The contours are at 30, 80, 160, 250 $\mu$Jy/b.a. Vectors of length of 
$1\arcsec$ correspond to a polarization degree of 1\%. The galaxy's centre is marked 
by a cross, while a triangle marks the region selected for energy density 
calculation (see Sect.~\ref{s:ongoing}). Two lines mark positions of performed slices
(Sect.~\ref{s:wave}).
}
\label{f:radio36}

\end{minipage}
\end{figure*}

The star formation traced by the ultraviolet emission of NGC\,4254 
observed by the OM through the UVW1 and UVW2 filters are shown as  
overlays in Fig.~\ref{radio21} (right panel) and Fig.~\ref{f:radio36} (top-right 
panel), respectively. They resemble the H$\alpha$ 
image (compare Fig.~\ref{xray}) but involve stars of intermediate masses 
and of ages of about 100\,Myr, the distribution of which is less clumped and even 
better follows the outline of spiral arms. 

\section{Results}
\label{s:results}


\subsection{Total radio intensity}
\label{s:total}
\begin{figure}

\resizebox{\hsize}{!}{\includegraphics{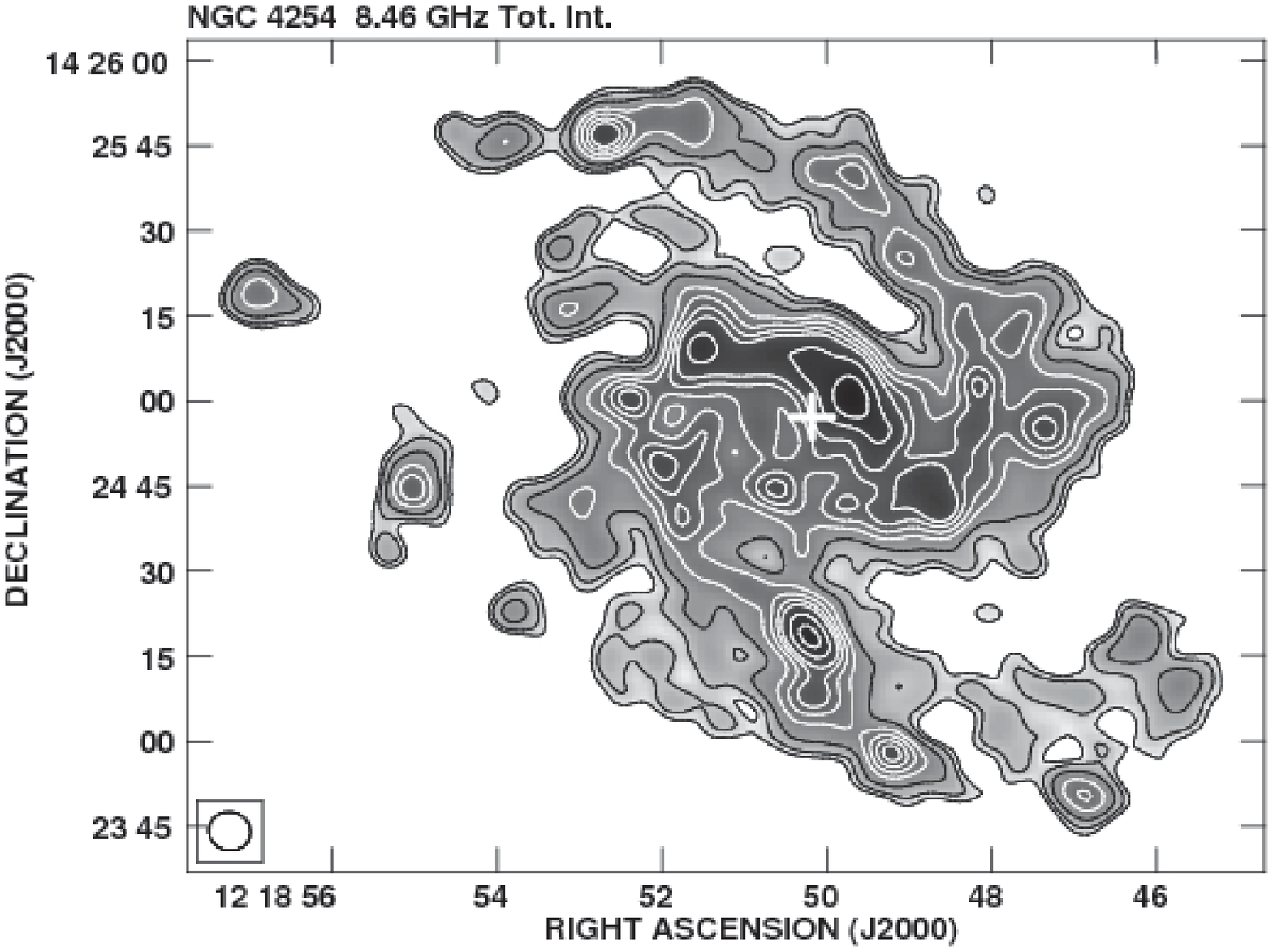}}

\caption{
The brightest structures in total radio intensity (greyscale and contours) of NGC\,4254 
at 8.46\,GHz observed at the highest resolution of $7\farcs 5$. The contours are at: 120, 
140, 180, 240, 300, 350, 400, 500, 600 $\mu$Jy/b.a. The cross denotes the position of 
the optical centre of the galaxy.
} 
\label{f:tparms}
\end{figure}

Our combined VLA and Effelsberg map of total radio intensity of NGC\,4254 at 
8.46\,GHz (Fig.~\ref{f:radio36}, top--left) shows a steep gradient of emission 
in the southern disk, where the long SW spiral arm is located 
(cf. Fig.~\ref{f:arms}). In the northern part of the disk the radio emission 
is much more diffuse and extends 1.5 times further from the centre than in the 
southern one. The same kind of N--S asymmetry is revealed by the optical B image 
(Fig.~\ref{f:radio36}, top--left in greyscale). The global N--S asymmetry 
is further confirmed by a more sensitive radio map at 4.86\,GHz 
(Fig.~\ref{f:radio36}, bottom--left). The asymmetric radio 
pattern is in this case more diffuse and extends roughly 
20\arcmin\ (1.6\,kpc) further out than at 8.46\,GHz.

Beyond the two brightest peaks in the radio images, which are background sources
(see Table~\ref{t:flux} for details), the galaxy's centre alone is the 
strongest source on the radio maps.
Contrary to the optical emission, which shows here numerous concentrations of
star formation not forming any particular pattern, the radio emission is elongated 
and hosts a bar--like feature, from which two arms (SW and NE) emerge. This is best
visible in our Fig.~\ref{f:tparms} of the central part of the disk with the highest 
resolution of $7\farcs 5$ (600\,pc). The two radio arms form a clear and symmetric 
spiral pattern. They extend from the bar ends to the north and south and then 
tightly wind around the disk. Similar features can be seen in the 
integrated CO emission (Sofue et al. \cite{sofue}), but not in the other ISM 
species.

Our 1.43\,GHz total intensity map with 17\arcsec\ HPBW resolution 
(Fig.~\ref{radio21}) reveals a very extensive diffuse radio envelope of NGC\,4254. 
The envelope is particularly stretched towards N, where the radio emission 
extends about $1\arcmin$ (more than 4\,kpc) further away than the optical 
emission in the disk. However, the radio envelope is 
surprisingly well traced by the \ion{H}{i} emission (shown in greyscale in 
Fig.~\ref{radio21}) -- not only the northern part but also  
two odd radio extensions to the NE (RA=$12^{\mathrm{h}} 19^{\mathrm{m}}	 
2^{\mathrm{s}}$, Dec=$14\degr 27\arcmin 50\arcsec$) and to the SW 
(RA=$12^{\mathrm{h}} 18^{\mathrm{m}} 37^{\mathrm{s}}$, Dec=$14\degr 
24\arcmin 0\arcsec$) have \ion{H}{i} counterparts. In the 
north, the \ion{H}{i} gas extends further than the radio emission. 
Also two \ion{H}{i} blobs in the south (around RA=$12^{\mathrm{h}} 18^{\mathrm{m}}
57.5^{\mathrm{s}}$, Dec=$14\degr 22\arcmin 40\arcsec$) 
have no radio counterparts. Both structures have \ion{H}{i} radial velocities 
which differ by about 100\,km\,s$^{-1}$ from the adjacent gas in the disk 
(see Fig.~\ref{f:hivelo}). They could be related to some 
external perturbing agent, as was suggested by Phookun et al. 
(\cite{phookun93}). 

We integrated the radio total emission over the galaxy. Due to different sizes 
of the VLA antenna primary beam for different frequencies, we applied 
the integration area extending up to about $3\farcm5$ in radius at 8.46\,GHz, 
and up to $5\arcmin$ at both 4.86\,GHz and 1.43\,GHz.  
The total radio fluxes at 4.86\,GHz and 8.46\,GHz are $162\pm6$\,mJy 
and $102\pm5$\,mJy, respectively. They are in a good agreement with the earlier 
works (e.g. Soida et al. \cite{soida}). The total integrated flux at 1.43\,GHz is 
$512\pm10$~mJy. This value is larger than that measured from the NVSS survey 
(418~mJy), due to the 10$\times$ better sensitivity of our present map. 
It is smaller than the single dish measurements (see Soida et al. 
\cite{soida}) most likely because of contamination from background sources in 
the previous low--resolution data. All background sources which can 
affect the integrated fluxes of NGC\,4254 are presented in Table~\ref{t:flux}. In all 
subsequent analyses the emission from background sources is subtracted.

\subsection{Polarized emission}
\label{s:polar}

The most characteristic feature of polarized radio emission at the high 
frequency of 8.46\,GHz is a strong ridge with a double peak in the 
southern part of the galaxy (Fig.~\ref{f:radio36},top--right). It shows a 
steep decrease of emission outwards from the disk. The ridge is clearly 
displaced by about 15\arcsec\ (1.2\,kpc) from the SW 
spiral arm and is located {\em outside} of it, on the downstream side 
of the density wave. The ridge has no counterpart in the total radio emission.

\begin{table}
\caption[]{Total radio fluxes of NGC\,4254 including the background sources (in mJy) and 
fluxes of the background sources alone at the three observed frequencies.
The integrated region was about $3\farcm5$ in radius at 8.46\,GHz and 
$5\arcmin$ at 4.86\,GHz and 1.43\,GHz.
}
\label{t:flux}
\centering
\begin{tabular}{lllll}
\hline\hline
RA$_{{\rm J}2000}$ & Dec$_{{\rm J}2000}$  & 8.46\,GHz & 4.86\,GHz & 1.43\,GHz \\
\hline
Total flux  &              & $102\pm5$     & $162\pm 6$   & $512\pm 19$\\
12 18 55.06 & 14 28 52.0   & n/a          & $1.1\pm 0.1$ & $5.7\pm 0.1$\\
12 18 45.70 & 14 26 44.6   & $4.0\pm 0.1$ & $7.1\pm 0.1$ & $22.4\pm 0.1$\\
12 18 42.15 & 14 23 07.0   & $1.8\pm 0.1$ & $2.1\pm 0.1$ & $4.0\pm 0.1$\\
\hline 
\end{tabular}
\end{table}


\begin{figure*} [t]

\centering
\includegraphics[width=8.9cm]{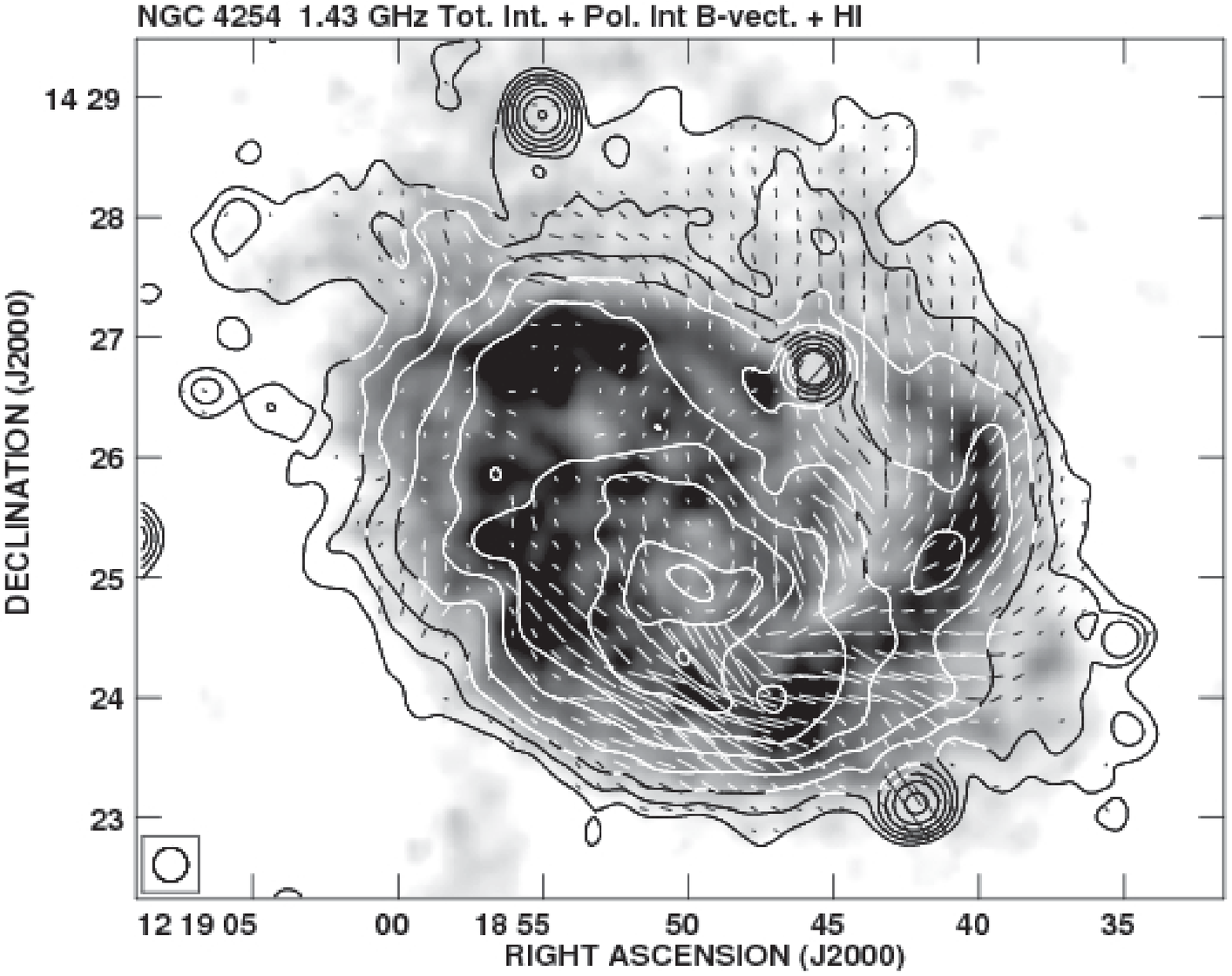}
\includegraphics[width=8.9cm]{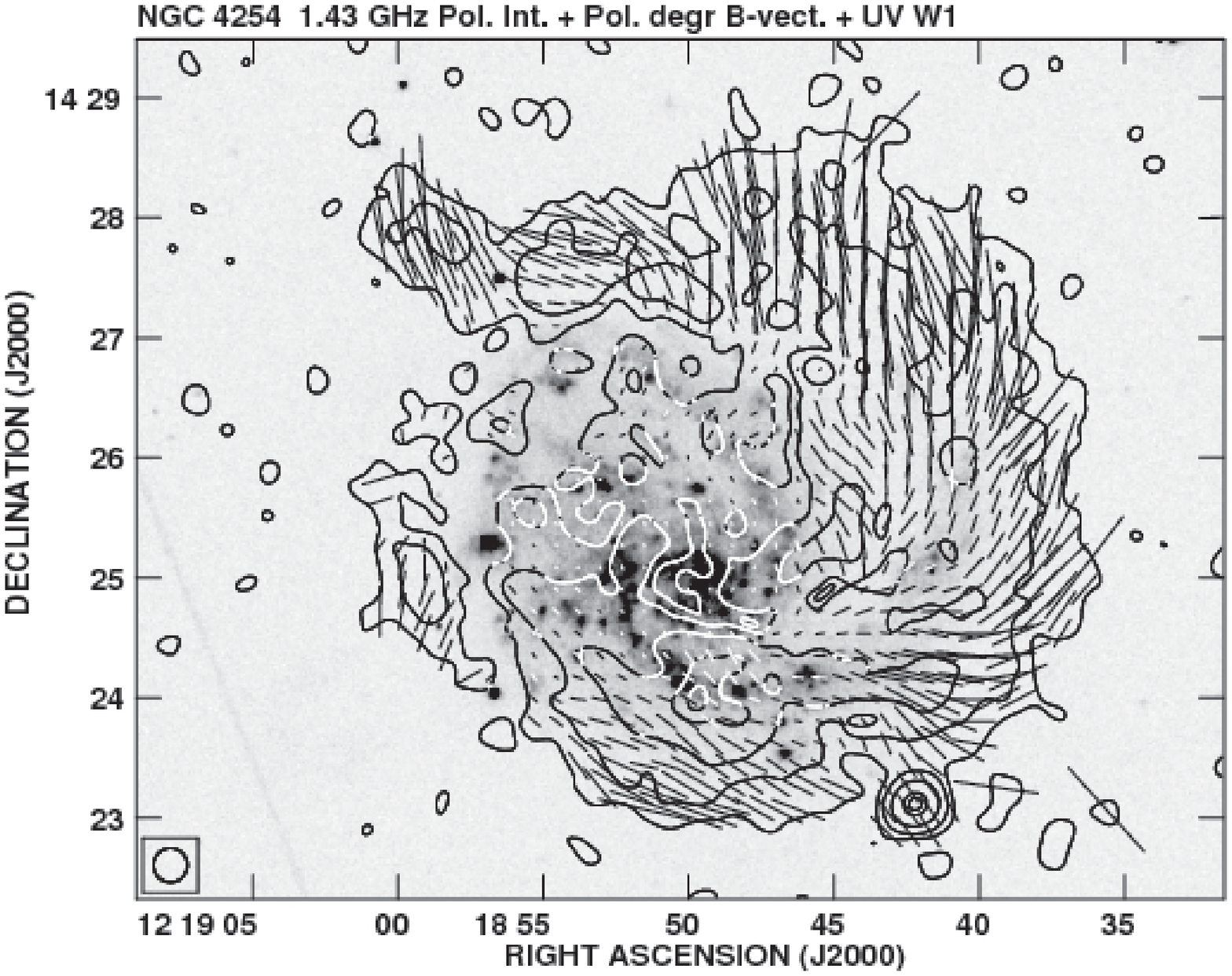}

\caption{
Left: The total intensity map of NGC\,4254 at 1.43\,GHz in contours with 
$\vec{B}$--vectors (not corrected for Faraday rotation) overlaid on the 
\ion{H}{i} line emission. The contours are 
0.06, 0.16, 0.32, 0.64, 1.3, 2.6, 5.1, 8.0, 10.2 mJy/b.a.  A vector of 
length of $1\arcsec$ corresponds to polarized intensity of 
$12.5\mu\mathrm{Jy/b.a.}$ Right: The polarized intensity contour map overlaid on 
the UVW1 filter map from our XMM satellite observations with $\vec{B}$--vectors 
of polarization percentage. The contours are 25, 70, 190, 450, 550 $\mu$Jy/b.a. 
A vector of length of $1\arcsec$ corresponds to a polarization degree of 
1\%. The beam size is $17\arcsec\times 17 \arcsec$.
}
\label{radio21}

\end{figure*}

At the base of the N arm (around RA=$12^{\rm h}18^{\rm m}45^{\rm s}$, 
Dec=$14\degr25\arcmin 0\arcsec$, Fig.~\ref{f:radio36}, top right), the highest 
polarization is found again outside the optical arm, like in the 
southern polarized ridge, on the downstream side of the density wave. In contrast, 
in the northern part of the same 
N optical arm (around RA=$12^{\rm h}18^{\rm m} 49^{\rm s}$, Dec=$14\degr26\arcmin 
0\arcsec$) the polarized emission is shifted {\em inwards}\ (to the east) from 
the spiral arm, thus to the upstream side of the density wave. Closer to 
the galaxy's centre, another strong elongated polarized feature is 
interlaced with the NE and SW spiral arms (around RA=$12^{\rm h}18^{\rm m}
51^{\rm s}$, Dec=$14\degr 25\arcmin 45\arcsec$). 

At 4.86\,GHz with higher sensitivity to extended structures, the polarized 
intensity covers almost the full optical extent of the galaxy (Fig.~\ref{f:radio36}, 
bottom right). Opposite to the southern polarized ridge the northern 
galaxy portion is filled with much more diffuse emission without any strong polarized 
maxima. This global N--S asymmetry in intensity is about 6, which is larger 
than either in total radio emission (about 4) or in optical, H$\alpha$ and \ion{H}{i} ones. 
This rises the question if the strong southern polarized ridge can result from 
the same perturbing agent as that causing the N--S asymmetry in other ISM components 
(see Sect.~\ref{s:puzzle}).

The $\vec{B}$--vectors of the ordered magnetic field\footnote{The ordered (regular) 
magnetic field observed in the polarized radio emission can consist of unidirectional 
(coherent) field and non-unidirectional anisotropic one. Faraday rotation is only sensitive 
to the coherent field component.} in NGC\,4254 form quite a smooth 
spiral--like pattern filling coherently the very broad (optically very 
faint) interarm region between the SW and N spiral arms (Fig.~\ref{f:radio36}, 
bottom left). As the Faraday rotation measure 
in NGC\,4254 only occasionally exceeds 100\,rad\,m$^{-2}$ (Paper II), 
the observed directions of the magnetic field at 8.46\,GHz and 4.86\,GHz are close 
to the internal ones (the differences are up to $7\degr$ and $23\degr$, respectively. 
The pitch angle of the observed magnetic field changes considerably throughout 
the galaxy from zero in the southern polarized ridge to about $40\degr$ 
in the northern part. However, the orientation of magnetic field is 
always close (within $20\degr$) to the direction of the nearby optical spiral arm, 
indicating that it is connected to the direction of the density wave. 

The galaxy's integrated polarized flux at 4.86\,GHz is $21\pm2$\,mJy,
which gives the mean polarization degree $p$ of $13.5\pm0.5\%$. The southern 
ridge of NGC\,4254 has a polarized flux of $3.1\pm0.3\,{\rm mJy}$, 
constituting about 15\% of the galaxy's total polarized emission. The mean degree of 
polarization in the ridge is about $23.8\pm0.5\%$, almost two times larger than 
the average over the whole galaxy. Without the ridge, the mean polarized fraction 
(in the rest of the galaxy) is 12.5\%, hence the ridge enhances the galaxy mean 
polarized fraction by only one percent. The highest degree of polarization 
(up to some 40\%) is found locally in the southern ridge and in the western interarm 
region (between the SW and N arms). 

Our low--frequency 1.43\,GHz data allow for a study of Faraday 
effects. At this frequency the Faraday rotation measure of 100\,rad\,m$^{-1}$ 
corresponds to the rotation of the magnetic field vectors by $253\degr$. Indeed, 
except the outer parts of the SW arm the orientations of the magnetic field seen in 
Fig.~\ref{radio21} are unrelated to those observed at higher 
frequencies (Fig.~\ref{f:radio36}, left). In contrast to the higher frequency maps, the 1.43\,GHz polarized 
intensity map (Fig.~\ref{radio21}, right) shows the strongly depolarized galaxy's 
centre ($p<1\%$) and the NE portion of the disk 
($p<3\%$), where patchy spiral arms are visible in 
the optical image. The western part is less depolarized, especially 
in the vast western interarm region ($p\approx 13\%$) and outside of 
star--forming regions along the SW optical arm ($p\approx 10\%$). This 
gives a strongly asymmetric distribution of 
1.43\,GHz polarized emission with respect both to the galactic major axis 
and to the location of the optically bright SW spiral arm.

The 1.43\,GHz data show another interesting feature: the reported extended 
envelope in the total radio intensity (Sect.~\ref{s:total}) turns out to be 
polarized, forming an  unclosed ring--like structure. 
The high polarization degree of $p\approx30-40\%$ indicates the presence of 
an ordered magnetic field extending further than the galactic gas traced by 
optical images. Similar polarized envelopes at 21\,cm were observed so far in M83 
(Neininger et al. \cite{neininger}) and NGC\,6946 (Beck\ \cite{beck07}). 
The polarized envelope of NGC\,4254 is the first one observed in a cluster spiral. 

\subsection{X--ray emission observed with XMM--Newton}
\label{s:xrays}

\begin{figure} [t]

\resizebox{\hsize}{!}{\includegraphics{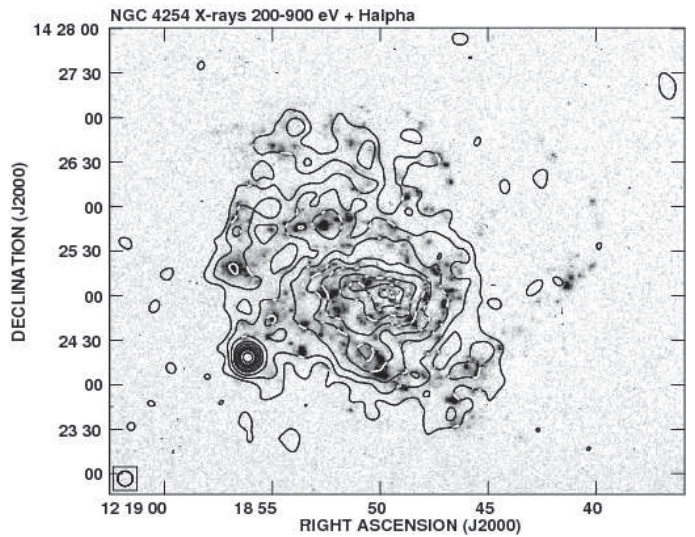}}
\caption{X--ray XMM--Newton emission of NGC\,4254 as seen by the combined EPIC cameras
smoothed to a resolution of $10\arcsec$.
Contours of the soft (200--900\,eV) emission are at (2.5, 4, 5.7, 8, 11.3, 16, 22.6, 32, 45.3, 64) 
$\times$ rms noise above the background and overlaid on the H$\alpha$ image of NGC\,4254 
(from Knapen et al. \cite{knapen04}). 
}
\label{xray}
\end{figure}

Our XMM--Newton data show the most extended emission of NGC\,4254 in soft energy 
bands (200--900\,eV, Fig.~\ref{xray}). The X--rays do not reveal 
either any feature outside the optical disk or any association with the southern 
strong polarized ridge. 

The X--ray emission is asymmetric: it is weak and diffuse in the
north and stronger and with a steeper gradient in the south. 
This pattern is similar to those seen in optical and H$\alpha$ (Fig.~\ref{xray}) 
and total radio images (Sect.~\ref{s:total}). Almost all X--ray features are associated with 
\ion{H}{ii} regions, they are particularly strong in the central part of 
the disk and in the regions located in the NW and NE spiral arms within 1\arcmin\ 
from the core. The central X--ray peak 
also corresponds well (in all energy bands) to the 
maxima in the optical and radio emission (the strongest peak at 
RA=$12^{\rm h}18^{\rm m} 56^{\rm s}$, Dec=$14\degr 24\arcmin 23\arcsec$ 
has no counterpart in other wavebands and is probably due to an unrelated 
source). The globally asymmetrical X--ray morphology and its correspondence 
with H$\alpha$ distribution implies that the hot gas is closely connected to 
the star--forming activity. Both 
disturbed morphologies resemble the observed total radio intensity at 8.46 and 4.85\,GHz 
(Sect.~\ref{s:total}) and are likely caused by the same perturbing agent. 

We do not find any X--ray background emission further out 
of the galactic disk out to $10\arcmin$ radius. This suggests that either NGC\,4254 
is situated outside the well--known hot envelope of the Virgo Cluster (B\"ohringer et al. 
\cite{bohringer}) or the intergalactic medium is so tenuous that it cannot 
be directly detected even in our sensitive observations.  

\section{Discussion}
\label{discussion}

\subsection{A brief comparison with other galaxies}
\label{s:compar}

Among all spiral non--barred galaxies mentioned in the recent reviews on 
galactic magnetism (Beck \cite{beck05}, Vall\'ee \cite{vallee04}, Widrow 
\cite{widrow02}) only about 10 have 1\,kpc--scale radio polarimetric 
observations with enough sensitivity to cover almost the whole stellar disk. 
Out of those only four (M31, M51, M83, NGC\,6946) have such high quality 
observations at three frequencies, like NGC\,4254. None of them is a cluster 
member. In comparing the radio properties of NGC\,4254 to other galaxies we found 
that the phenomenon of polarized ridges shifted from optical spiral arms is 
a most peculiar one. To some extent the ridges resemble the observed magnetic arms in 
the non--interacting non--cluster spirals: IC342 (Krause et al. \cite{krause89},
Krause \cite{krause93}); and NGC\,6946 (Beck \& Hoernes 
\cite{beckhoernes}). The high degree of polarization in the southern ridge 
in NGC\,4254 (up to $p=40\%$ with a mean of 24\%, Sect.~\ref{s:polar}) is also 
similar to the bright magnetic arms in NGC\,6946 (40\% and 28\%, respectively, Beck 
\& Hoernes \cite{beckhoernes}, Beck \cite{beck07}).
However, the southern polarized ridge in NGC\,4254 is a very special feature. In 
typical grand--design spirals like NGC\,6946, and also in some portions of 
NGC\,4254, ordered fields are strongest at the locations of prominent dust 
lanes on the inner edge of gaseous spiral arms (Beck \cite{beck05}). 
But the strong southern ridge in NGC\,4254 is located downstream the density wave 
at the edge of the galactic disk (Fig.~\ref{f:radio36}, top--right), which can suggest 
some unknown, possibly external, process.

The southern ridge also drives a strong global N--S asymmetry in the polarized intensity. 
Such asymmetric polarization patterns have been observed to date only for some 
interacting galaxies as NGC\,3627 (Soida et al. \cite{soida01}) or NGC\,2276 
(Hummel \& Beck \cite{hummel95}). This implies some external interaction also in the 
case of NGC\,4254. However in those galaxies the polarized signal is not shifted out 
of the bright optical disk.

In the most western portion of the disk the polarized emission in NGC\,4254 
follows the optical spiral arm (see also Sect.~\ref{s:wave}). This resembles outer 
southern and southwestern spiral arms in M\,51 (Patrikeev et al. \cite{patrikeev06}).  
Ordered fields coinciding with the outer optical arms were also reported for M83 
and NGC\,2997 (Beck \cite{beck05}). In short, the polarized properties of NGC\,4254 are 
inhomogeneous and, besides the southern ridge, show a mixture of structures 
observed in other galaxies.

Concerning the X--ray emission, we notice that the only other 
spiral--like galaxy in the Virgo cluster for which high sensitivity X--ray data 
are published is NGC\,4438 (Machacek et al. \cite{machacek04}). This edge--on 
galaxy has a highly disturbed morphology in the stellar content, which is 
followed by H$\alpha$ and X--ray emissions displaced up to 10\,kpc from the disk's plane. 
The galaxy probably suffered from a strong off--centre collision with 
the nearby elliptical galaxy NGC\,4435 about 100\,Myr ago (Machacek et al. 
\cite{machacek04}). The X--ray emission in NGC\,4254 also shows a close connection 
with the perturbed stellar disk and H$\alpha$ gas, but the galaxy is not "damaged" 
as NGC\,4438, which may suggest a much weaker interaction during a possible encounter 
(see Sect.~\ref{s:tidal}).

Another process that could perturb NGC\,4254 and its X--ray emission is ram 
pressure due to the motion of the galaxy through the hot ICM. An example of 
such a process at work can be seen in the case of NGC\,2276, a member of the 
NGC\,2300 group of galaxies. The Chandra X--ray observations reveal a bow shock 
feature along the western galactic disk with enhanced density and temperature 
of hot gas (Rasmussen et al. \cite{rasmussen06}). This is also the location 
of shocked and compressed magnetized plasma, as the radio total and polarized emissions 
peak here (Hummel \& Beck \cite{hummel95}). Furthermore, a tail of diffuse hot gas extends out 
of the galaxy on the opposite disk side. None of the X--ray features connected with 
the ram pressure forces in NGC\,2276 (the leading bow shock and the hot tail) are 
present in the X--ray image of NGC\,4254 (Fig.~\ref{xray}), which makes a point against 
strong ram pressure effects in this case.

Therefore, it is not easy to construct a single interaction scenario for NGC\,4254. 
Some of its distinct observed properties are either unknown among 
galaxies (the southern polarized ridge) or can vaguely indicate weak 
environmental effects of some kind (the X--ray emission). A more detailed 
analysis of this problem is presented below. We start 
with the discussion of the radio spectral index and the separation of thermal and 
nonthermal radio emission.

\subsection{Spectral index distribution}
\label{s:spectral}

\begin{figure}

\resizebox{\hsize}{!}{\includegraphics{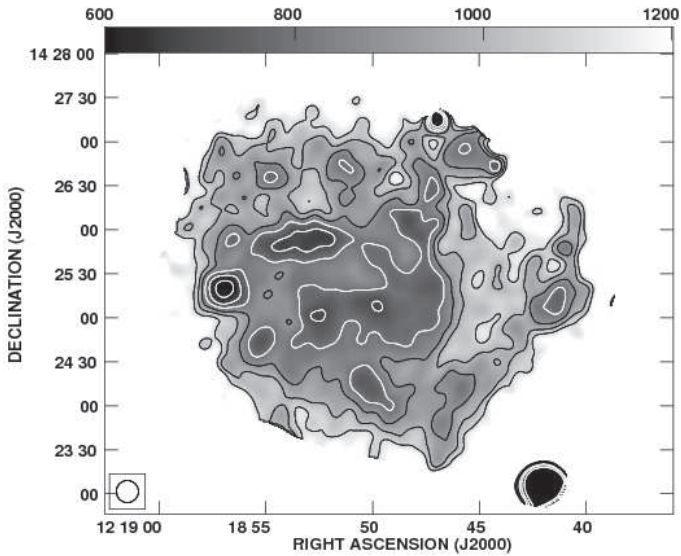}}

\caption{
Radio spectral index distribution between 4.86 and 1.43\,GHz in NGC\,4254 (contours 
and greyscale). Both maps of total intensity were convolved to a common beam of 
$15\arcsec$. The contours are: 0.6, 0.7, 0.8, 
0.9,  1.0, 1.1.} 
\label{spix}
\end{figure}

The spectral index distribution in NGC\,4254 (Fig.~\ref{spix}) has been 
determined for each map pixel according to a power law 
S$_{\nu}\propto\nu^{-\alpha}$ using the values of total intensity at 
1.43\,GHz and 4.86\,GHz. 
At least half of the galactic regions have a steep radio spectrum with $\alpha\simeq 
0.8$, which is also the mean galactic spectral index, typical for late--type spirals 
(e.g. Niklas et al. \cite{niklas97b}). The spectrum flattens in the regions of 
spiral arms. Within the heavy concentrations of star formation visible in H$\alpha$ 
emission it reaches a slope of about 0.7, while in the nuclear region it is just 0.64. 
The flattest spectrum of 0.55 is observed at the position of the strongest 
\ion{H}{ii} region at the tip of NE arm (RA=$12^{\rm h}18^{\rm m}57^{\rm s}$, 
Dec=$14\degr 25\arcmin 19\arcsec$). Its radio and X--ray emission is present 
but not too strong, which may indicate some very recent star--forming activity. Outside of 
spiral arms the spectrum steepens, achieving the slope 
of about 1.2 at the disk's outskirts. In the characteristic large interarm space between the 
SW and N arms the spectral index varies from about 1.0 to 1.1. 

There is no global N--S asymmetry in the spectral index distribution as that 
observed in total radio intensity. There is no evidence for cosmic rays (CRs) 
acceleration or enhanced ageing of CRs in the southern polarized ridge, which 
would indicate a shock or strongly compressed magnetic fields; the spectral 
index of about 1.0 in this area is similar to that of typical interarm regions in the NE 
disk.

\subsection{Thermal and nonthermal components}
\label{s:separation}


\begin{figure*} [t]
\centering

\includegraphics[width=8.9cm]{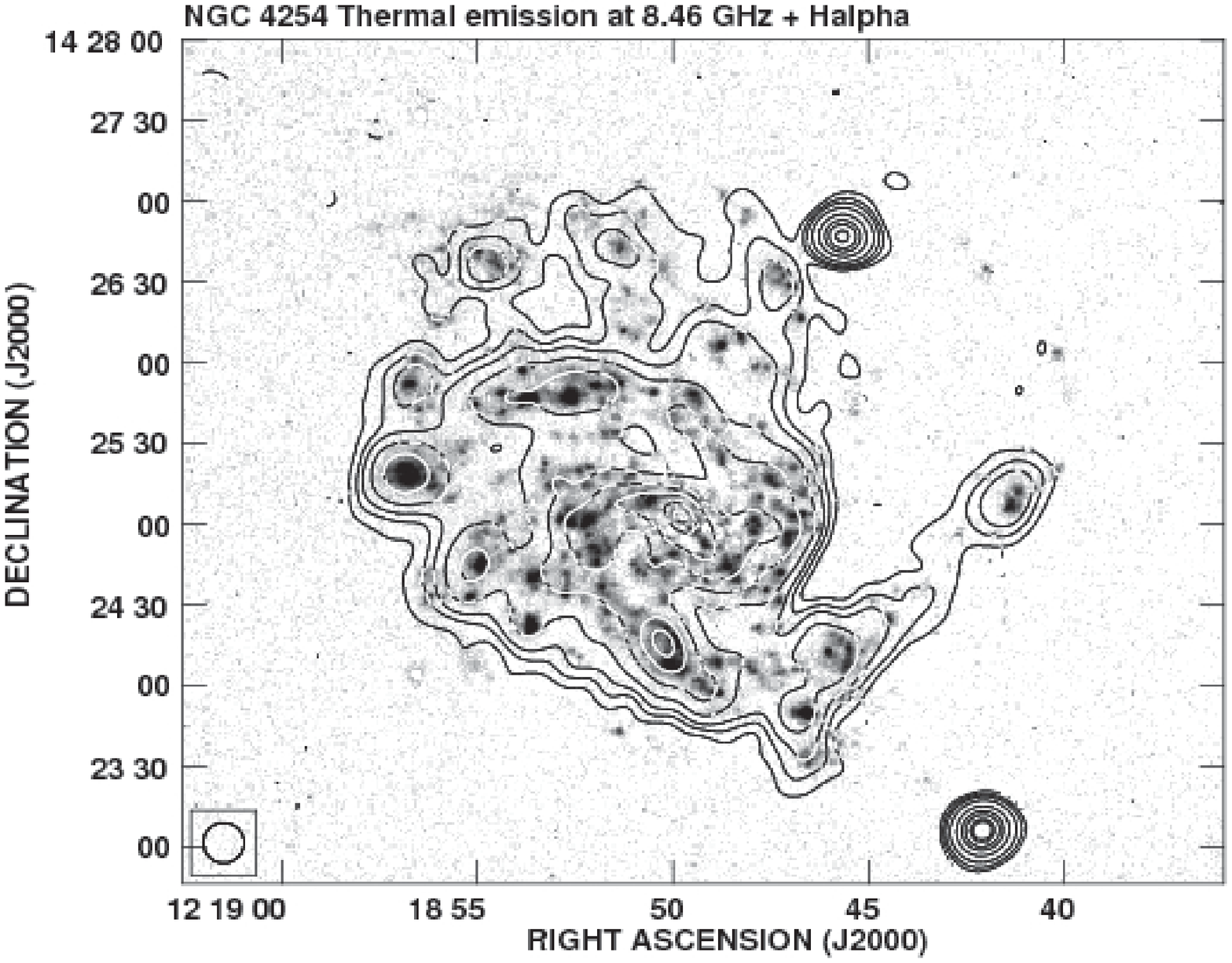}
\includegraphics[width=8.9cm]{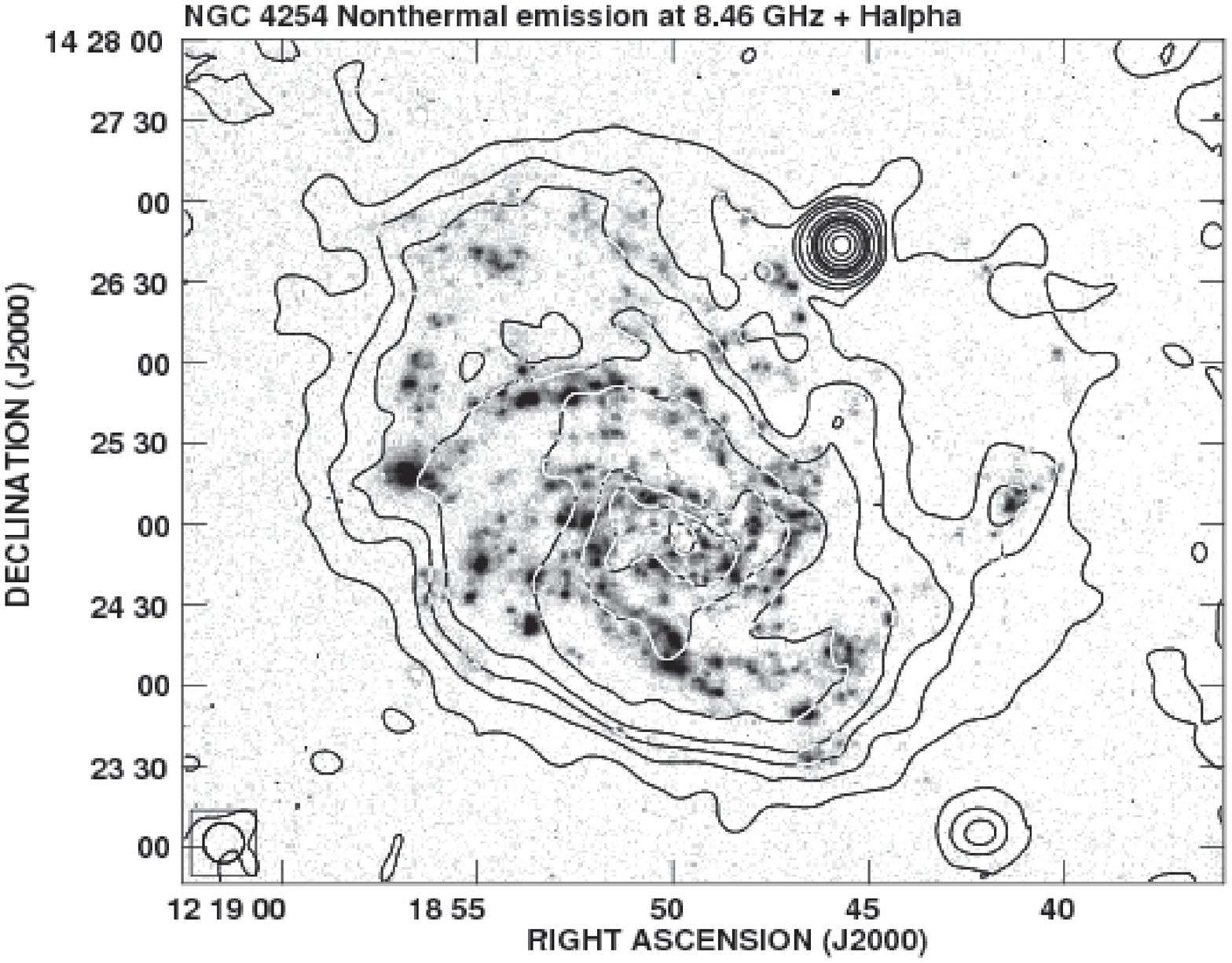}

\caption{
Thermal (left panel) and nonthermal (right panel) radio intensity maps of NGC\,4254 
at 8.46\,GHz overlaid on the H$\alpha$ image. 
Contours are: left panel: 0.06, 0.16, 0.32, 1.2, 2.5, mJy/b.a.; right panel: 
25, 70, 190, 450, 559 $\mu$Jy/b.a. 
The beam size is $15\arcsec\times 15 \arcsec $.
}
\label{th}

\end{figure*}

The distributions of synchrotron and thermal free--free radiation in NGC\,4254 
can be separated by least--square fits of thermal fraction in each of the map--points 
of observed radio emission simultaneously at three frequencies. We assume 
that the nonthermal spectral index $\alpha_{nth}$ is constant within the 
frequency range 1.43--8.46\,GHz and throughout the galaxy (see e.g. Ehle \& 
Beck \cite{ehle93}). The synchrotron spectrum is flatter in the spiral arms, 
which leads to overestimating the thermal intensity and underestimate the nonthermal
one (Tabatabaei et al. \cite{taba07b}). We performed the fitting procedure for various 
values of $\alpha_{nth}$ from 0.95 to 1.1 and compared the derived distribution 
of radio thermal emission with the H$\alpha$ map. The best correspondence in 
the interarm regions, where possible H$\alpha$ dust attenuation is minimized, 
was achieved for $\alpha_{nth}\approx 1.0 \pm0.05$. We checked this value by 
two other methods. In the first approach we took the integrated 
fluxes at three frequencies and separated the global thermal and nonthermal 
components in a way similar to Niklas et al. (\cite{niklas97b}). This yields 
$\alpha_{nth}=1.0$. Next, we estimated $\alpha_{nth}$ using our spectral index 
map (Fig.~\ref{spix}) taking the mean value in the large interarm region between SW 
and N arms being free from thermal emission. This gave $\alpha_{nth}=1.05$. 
Since all methods give consistent results we use finally $\alpha_{nth}=1.0$.

A close similarity of the obtained radio thermal emission and H$\alpha$ emitting 
gas (Fig.~\ref{th}) confirms that the applied procedure cannot be significantly influenced by systematic 
effects. The thermal fraction at 8.46\,GHz is typically up to about 25\% in 
the interarm space and $35\%-55\%$ within the spiral arms. 
The mean thermal fraction over the whole galaxy of 20\% at 
8.46\,GHz is lower than the mean fraction of $30\%\pm 4\%$ calculated 
for a large sample of disk galaxies by Niklas et al. (\cite{niklas97b}). 
However, systematically lower thermal fractions were found by Niklas 
(\cite{niklas95}) in galaxies with higher star formation rates, which is
the case for NGC\,4254 (see Sect.~\ref{s:sfr}).

The distribution of nonthermal emission is remarkably smooth (Fig.~\ref{th}), 
partly due to CR diffusion process. However, it reaches 
up to different radii in different galaxy regions: about $2-2.5\,{\rm kpc}$ in 
the southern polarized ridge and up to 4\,kpc in the western interarm space. 
If the radial extent of magnetic field is similar in those regions, this is 
likely to be the result of different diffusion coefficients for CRs 
propagating at various orientations to the mean magnetic field (see e.g. 
Giacalone \& Jokopii \cite{giacalone99}). The striking result is the lack 
of any ridge of total synchrotron radiation to the south from the SW arm, 
where the {\em polarized} synchrotron emission has a global maximum within the whole 
galaxy disk (Sect.~\ref{s:polar}). We explain this by 1) a relatively small 
contribution of polarized emission from ordered magnetic fields to the total 
synchrotron radiation here (polarization 
degree of about $p=0.24\%$), and 2) gradual increase 
of actual synchrotron emission towards the galaxy's centre due to higher density 
of CRs and random magnetic fields.

\subsection{Interrelations between gas phases}
\label{s:wave}

The interrelations between various ISM components in NGC\,4254 were analysed 
using a spectral decomposition of the related images in the broad domain of spatial 
scales (Frick et al. \cite{frick01}). Besides radio, X--ray, and optical data, we 
included the $24\mu{\rm m}$ mid--infrared (MIR) map from the Spitzer survey of SINGS galaxies 
(Kennicutt et al. \cite{kennicutt03}), revealing dust and PAH emission heated 
by young stars and IRAC $3.6\,\mu{\rm m}$ (near--infrared, NIR) map dominated by radiation from old 
stellar atmospheres. All maps were convolved to the same resolution of $15\arcsec$. 
The brightest point-like sources were subtracted from the maps.
We applied the {\em Pet Hat} (PH) isotropic wavelet to derive the wavelet 
cross--correlation coefficient $r_w$ at various spatial scales. 

\begin{figure}

\center
\includegraphics[width=8.8cm]{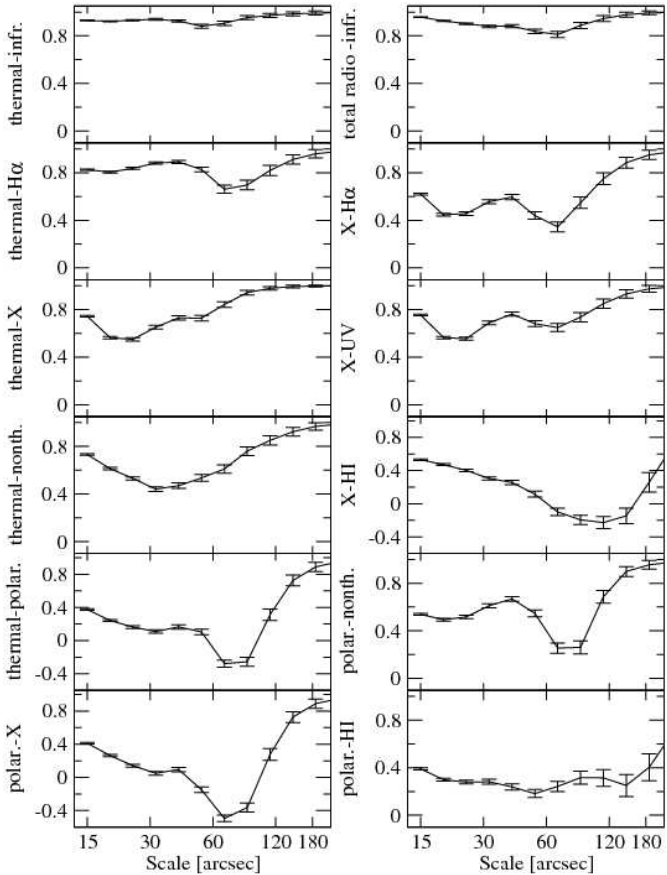}

\caption{
Wavelet (PH) cross--correlation coefficients for some ISM species as a function 
of scale. 
} 
\label{f:corel}
\end{figure}

A high cross--correlation ($r_w\approx0.8$) is observed between almost all
species at largest scales (Fig.~\ref{f:corel}), demonstrating the overall 
correspondence of their extended emission and hence the lack of spatial truncation 
in any gas phase. A lower correlation of hot ionized medium traced by our X--ray data 
with \ion{H}{i} ($r_w\approx0.3$) reflects an extended envelope of neutral hydrogen 
discussed in Sect.~\ref{s:polar}.

The highest wavelet--correlations are visible between the various species of warm 
ionized medium (WIM) dominated by thermal processes. The 
highest correlation ($r_w>0.88$) is found between the radio thermal and mid--infrared 
emissions at all spatial scales, and between the total radio emission and 
the mid--infrared one ($r_w>0.80$). Almost similarly high ($r_w\approx 0.80$) is the 
radio thermal--H$\alpha$ correlation. The synchrotron emission is also connected to 
the thermal emission, although with smaller correlation ($r_w>0.4$). 
Similar relations are observed in the grand--design spiral 
NGC\,6946 (Frick et al. \cite{frick01}), e.g. for the H$\alpha$--radio thermal emissions 
$r_w=0.8$ and for the radio thermal--nonthermal one $r_w=0.53$.
This indicates that the cluster medium does not affect spatial interrelations 
between various ISM components in NGC\,4254 and they remain quite similar to those in 
the non--cluster and non--interacting spiral NGC\,6946.

A completely different kind of relations is revealed by the polarized emission, which 
shows a decrease of correlation with other species at scales of about 1\arcmin\ to 
$1\farcm 5$. In the case of thermal and X--ray emission it even amounts to 
anticorrelation. Such anticorrelation was also observed in NGC\,6946 
($r_w=-0.1$ for the radio polarized and thermal emission, Frick et al. \cite{frick01}), 
but here it is even stronger ($r_w=-0.4$). 
It is obviously caused by a distinct displacement of the southern polarized ridge from thermal 
(warm and hot) gas located in the NW spiral arm, and by the other polarized features 
interlaced with optical arms (Sect.~\ref{s:polar}).

\begin{figure} 
\includegraphics[width=8.9cm]{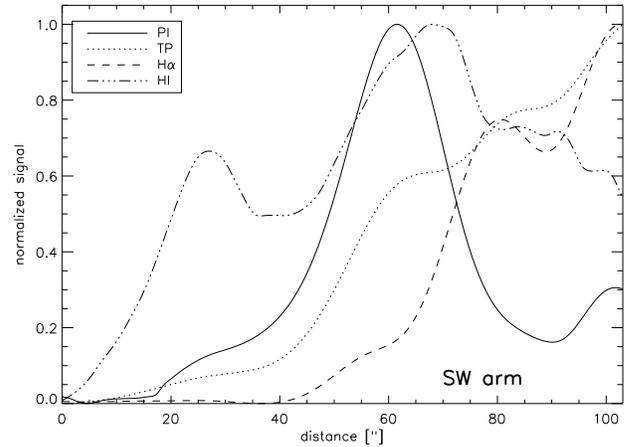}
\caption{The southern slice across the SW arm in polarized and total radio 
intensity at 4.86\,GHz, as well as H$\alpha$ and \ion{H}{i} emission in 15\arcsec\ 
resolution. The direction of the slice is from outside towards the disk centre 
(see Fig.~\ref{f:radio36}, bottom-right panel, for the exact position). The 
location of the SW optical spiral arm is also marked. 
} 
\label{f:SWslice}
\end{figure}
In order to further investigate the phenomenon of polarized emission in NGC\,4254, we 
performed a slice across the southern polarized ridge in distributions of various 
ISM components. The slice (presented in  (Fig.~\ref{f:SWslice}) shows clearly a 
displacement of polarized intensity 
with respect to the optical arm by about 20\arcsec\ (a maximum in the polarized 
intensity is located 60\arcsec\ from the slice's beginning, while the H$\alpha$ peaks 
at 80\arcsec). Moving towards the galaxy centre, the optical as well as the total radio 
emission increase, while the polarized emission drops off steadily, possibly due 
to an enhanced star--forming activity, and thus a larger ISM turbulence, which disrupts 
the ordered field. There is a correspondence of the maximum in polarized emission 
with the enhancement of \ion{H}{i} distribution, which also peaks outwards of the 
optical arm (at 70\arcsec). An additional rise in the \ion{H}{i} distribution can be seen 
40\arcsec\ further out of the polarized ridge, but neither polarization nor other gas tracers 
are found here. 

\begin{figure} 
\includegraphics[width=8.9cm]{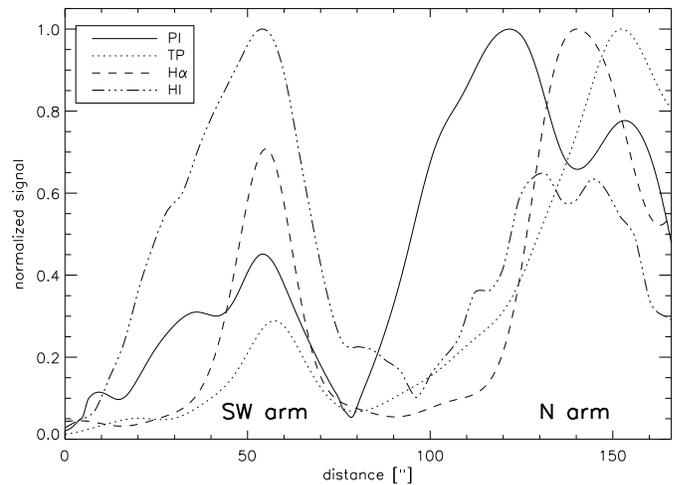}
\caption{The western slice across the interarm (N and SW) region. The details as in 
Fig.~\ref{f:SWslice}. The locations of optical spiral arms are marked.
} 
\label{f:interarmslice}
\end{figure}
Can accumulations of \ion{H}{i} gas alone be a driver for the enhancements of 
polarized emission in NGC\,4254? In order to test this idea we performed another slice 
through the western part of the galaxy across the N and SW arms (Fig.~\ref{f:interarmslice}).  
The \ion{H}{i} gas peaks now at the exact position of the optical SW arm (at 55\arcsec) 
and is associated with only a small peak in polarized 
emission (0.4 in normalized signal). However, the more than two times weaker \ion{H}{i} 
emission around the N arm is accompanied by the global maximum in polarized intensity 
(1.0 in the normalized signal at 120\arcsec). This is quite contrary to the southern 
part of the disk, where polarized and \ion{H}{i} maxima coincide.

It seems that the strongest polarized emission in NGC\,4254 is additionally enhanced 
outside the optical arms by some external process independent of \ion{H}{i} gas 
(see Sect.~\ref{s:puzzle}). Hence, the polarized emission behaves in a very different 
way than the other ISM species analysed and is of significance in any modelling of 
ISM evolution, providing us with independent information on the history of cosmic ray 
and the magnetic field evolution.

\subsection{Radio thermal emission as a star formation rate indicator}
\label{s:sfr}

One of the best measures of evolutionary state of a galaxy and a sensitive
indicator of some types of environmental interactions is SFR (e.g. 
Koopmann \& Kenney \cite{koopmann04}). Many SFR indicators
basing on luminosities from various spectral ranges were used
(e.g. Kennicutt \cite{kennicutt98}), but they all suffer from different 
flaws: the most common H$\alpha$--based SFR is severely 
contaminated by internal extinction, while the extinction--free, infrared--based SFR  
is affected by an unknown fraction of diffuse IR emission (cirrus 
component) and requires a high degree of dust opacity. Condon (\cite{condon92}) derived 
a commonly used radio nonthermal SFR indicator, basing on empirical 
calibration of radio emission by the supernova rate within the Milky Way,
which however may be not valid for all galaxies.

The most straightforward and accurate SFR indicator can be derived from the radio thermal 
emission, which is free from all the mentioned difficulties and, in addition, can 
also be applied within a single galaxy. The most severe problem with this approach is 
uncertainty of the thermal--nonthermal flux separation, which requires 
sensitive multi--frequency radio data. However, in the case of NGC\,4254 the 
separation procedure proved to be reliable and allows us to estimate  
SFR in the individual regions of NGC\,4254. For the first time we
confront the derived SFR {\em within} a spiral galaxy basing on thermal radio 
emission with the H$\alpha$--based SFR.

\begin{figure} [t]

\resizebox{\hsize}{!}{\includegraphics{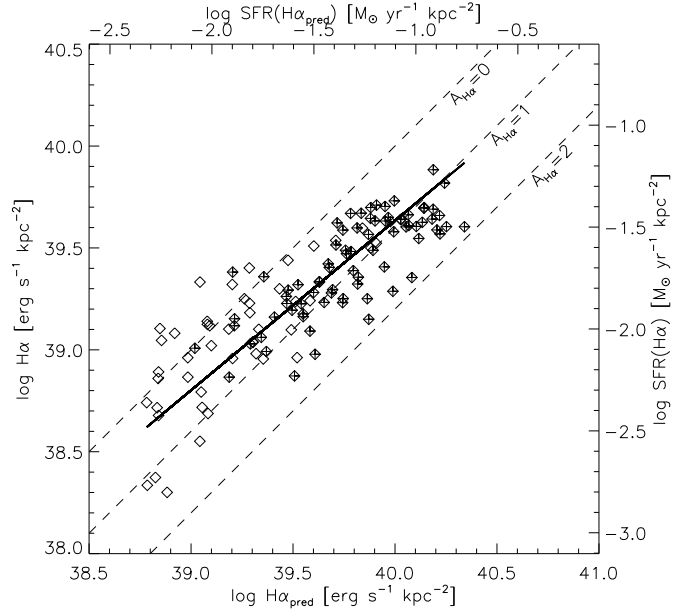}}

\caption{
The luminosities of H$\alpha$ emission and the predicted extinction--free 
H$\alpha_{pred}$ emission from the radio thermal data for the individual 
beam--separated regions of 
NGC\,4254. The luminosities of all regions are normalized to the area in 
kpc$^2$. The crosses indicate the regions in the inner part of the galaxy with 
higher signal--to--noise ratio. The solid line shows the best--fit linear model, 
while the dashed lines represent internal H$\alpha$  extinction values of 0, 1, and 
2 magnitudes. The corresponding SFRs are indicated on the top and right coordinate axes.
}
\label{f:sfr}
\end{figure}

We used the map of radio thermal emission at 8.46\,GHz and the map of 
H$\alpha$ emission from Knapen et al. (\cite{knapen04}) convolved to a common 
beam of 15$\arcsec$ HPBW. The H$\alpha$ map was corrected for baselevel undulations 
using the task IMSURFIT in IRAF package. The background constant level was then 
subtracted and the image calibrated with the scaling factor of 38.0926 
from Knapen et al. (\cite{knapen04}). We corrected the H$\alpha$ emission for  
foreground Milky Way extinction using $E(B-V)=0.039$ (from LEDA). With the 
adopted distance of 16.8\,Mpc to the galaxy we derived a total H$\alpha$ flux 
of $1.35 \times 10^{-11}$ erg\,s$^{-1}$, which is in good agreement with Koopmann 
et al. (\cite{koopmann01}). Both maps were decomposed into beam--separated 
regions to determine values of radio thermal and H$\alpha$ brightness. The 
radio thermal fluxes were then used to predict the extinction--free emission in 
the H$\alpha$ line, H$\alpha_{pred}$, according to the classical model of 
\ion{H}{ii} regions by Caplan \& Deharveng (\cite{caplan86}). Following this 
approach, we evaluated the predicted H$\alpha$ emission from the thermal flux 
$TH_\nu$ at the frequency $\nu$:
\begin{eqnarray}
    \left(\frac{ \mathrm{H}\alpha_{pred}}{\mathrm{erg\, s^{-1}\, cm^{-2}}}\right) &=&
3.5\times 10^{-7}\,\frac{N(H^+)}{N(H^+)+N(He^+)} \cdot \\
\nonumber & \cdot & \left(\frac{T_e}{\mathrm{K}}\right)^{-0.57}
    \left(\frac{\nu}{\mathrm{Hz}}\right)^{-0.1}TH_\nu
\end{eqnarray}
For electron temperature we assumed a typical value of $T_e=10^4$\,K, for abundance 
ratio $N(H^+)/N(He^+)=0.08$, and for frequency $\nu=8.46$\,GHz. The results
weakly depend on temperature and He/H ratio: rising the 
temperature to $1.2\times 10^4$\,K, or the abundance ratio to 0.09 would change 
H$\alpha_{pred}$ values by 10\%. Both the observed H$\alpha$ and predicted 
H$\alpha_{pred}$ fluxes were converted to luminosities, adopting the distance of 16.8\,Mpc 
to NGC\,4254, and normalized to the area in kpc$^2$ in each independent region. These 
luminosities together with the corresponding SFRs derived 
from the extinction--free H$\alpha$ calibrator 
(Kennicutt \cite{kennicutt98}) are plotted in Fig.~\ref{f:sfr} 

The predicted extinction--free H$\alpha$ luminosities basing on the radio thermal emission 
are on average larger than the measured H$\alpha$ ones, as it could be expected. The mean 
internal extinction for the whole galaxy is about 0.8 mag, but
seems to depend on luminosity: for low luminosities of about 
$10^{39}$~erg~s$^{-1}$~kpc$^{-2}$ it is about 0.5 mag and 
systematically rises to about 1.0 mag at 2$\times10^{40}$~erg~s$^{-1}$~kpc$^{-2}$. 
This behaviour is quantified by a linear fit to the data, using the bisector 
method by Isobe et al. (\cite{isobe90}), giving:
\begin{equation}
\mathrm{log}(\mathrm{H}\alpha)=(6.3\pm2.2)+(0.83\pm0.06)\,\mathrm{log}(\mathrm{H}\alpha_{pred})
\end{equation}
or in other form:
\begin{equation}
A_{\mathrm{H}\alpha}=(0.42\pm0.15)\,\mathrm{log}(\mathrm{H}\alpha_{pred})-(18.8\pm5.5)
\end{equation}
The observed rise of extinction with H$\alpha$ emission can be explained 
by a larger amount of dust and CO gas within intense star--forming regions.
In order to check for possible influencing this relation by regions with lower 
signal--to--noise ratio, typically found in the outer parts of the galaxy, 
(indicated by crosses in Fig.~\ref{f:sfr}) we performed a similar fit 
without them. The results are very similar and the slope in the 
first relation above is 0.80~($\pm$0.05) in this case.

According to our findings, the mean SFR of NGC\,4254 as a whole is 
$0.026~\mathrm{M}_{\sun}~\mathrm{yr}^{-1}~\mathrm{kpc}^{-2}$ -- about 
3 times larger than in the nearby normal galaxies of similar Hubble types 
(see Kennicutt \cite{kennicutt98}). This result is contrary to the 
main trend among Virgo Cluster spirals, which show SFR reduced by a factor 
up to 2.5 as compared to isolated objects (Koopmann \& Kenney \cite{koopmann04}). 
The reduction of SFR was found in galaxies with spatially truncated 
H$\alpha$ distribution in the outer disk portions and was explained by a 
stripping process exerted by the ram pressure of hot ICM (see also 
Sect.~\ref{s:ongoing}). 

\begin{table}
\caption{Global SFRs for NGC\,4254, corrected (where necessary) for Milky 
Way extinction.
}
\label{t:sfr}
\centering
\begin{tabular}{ll}
\hline\hline\noalign{\smallskip}
Method   &  SFR \\
         &  [$M_{\sun}~yr^{-1}$]\\
\hline
Thermal radio $L_{H\alpha pred}=1.1 \times 10^{42}$ erg s$^{-1}$ & 8.4 \\
Total radio $L_{1.4}=1.62 \times 10^{29}$ erg s$^{-1}$ Hz$^{-1}$ & 9.6 \\ 
MIR $L_{24\mu m}=1.39 \times 10^{30}$ erg s$^{-1}$ Hz$^{-1}$ & 7.1 \\ 
H$\alpha$ $L_{H\alpha}=5.0 \times 10^{41}$ erg s$^{-1}$ & 4.0 \\
\hline
\end{tabular}
\end{table}

Finally, as a check for consistency, we compare in Table~\ref{t:sfr} 
the {\em global} SFR derived for NGC\,4254 using various indicators according to 
Kennicutt (\cite{kennicutt98}). We assume the Salpeter 
IMF in the range of 0.1--100\,$M_{\sun}$ in each case. The infrared estimation 
($7.1\, M_{\sun}\,\rm{yr}^{-1}$) and our radio thermal emission indicator 
($8.4\,M_{\sun}\,\rm{yr}^{-1}$) agree within the errors of 
calibration, which are of about 10\% and 5\% in the MIR and radio bands, respectively. 
The SFR value derived from the total radio 
luminosity at 1.43\,GHz exceeds by 35\% the one based on the infrared emission. 
The difference is physically relevant. At this 
frequency the total radio flux is largely dominated by synchrotron emission, 
which is nonlinearly related to SFR with an exponent larger than 1 
(Niklas \& Beck \cite{niklas97a}, see also next Section). This makes the 
nonthermal radio indicator less precise and subject to systematic bias.
The H$\alpha$--based SFR is underestimated (by a factor of about 2)
and without correction for dust extinction can lead to serious errors.

\subsection{Radio thermal --  nonthermal -- IR relations}
\label{s:fir}

The well known global correlation of the radio and infrared emission of galaxies 
(Helou et al. \cite{helou85}) was demonstrated to be present in our vicinity as well as in 
the distant Universe (Appleton et al. \cite{appleton04}). The underlying physical 
processes of this relation obviously act at scales much smaller than galaxy sizes, 
but their understanding is 
still incomplete. Taking the opportunity of Spitzer's recent high resolution  
observation of warm dust of NGC\,4254 at mid--infrared $24\,\mu$m we investigated the 
radio--infrared relation {\em within} this galaxy separately for 
the thermal and nonthermal radio components decomposed in 
Sect.~\ref{s:separation}. Performing such a study for the first time for a cluster galaxy is of 
a special interest, as the influence of the cluster environment on the radio--IR 
relation is still unknown and NGC\,4254 shows distinctly a disturbed morphology 
in all spectral bands as well as enhanced processes of star formation (Sect.~\ref{s:sfr}). 

\begin{figure*} 
\center
\includegraphics[width=6.05cm]{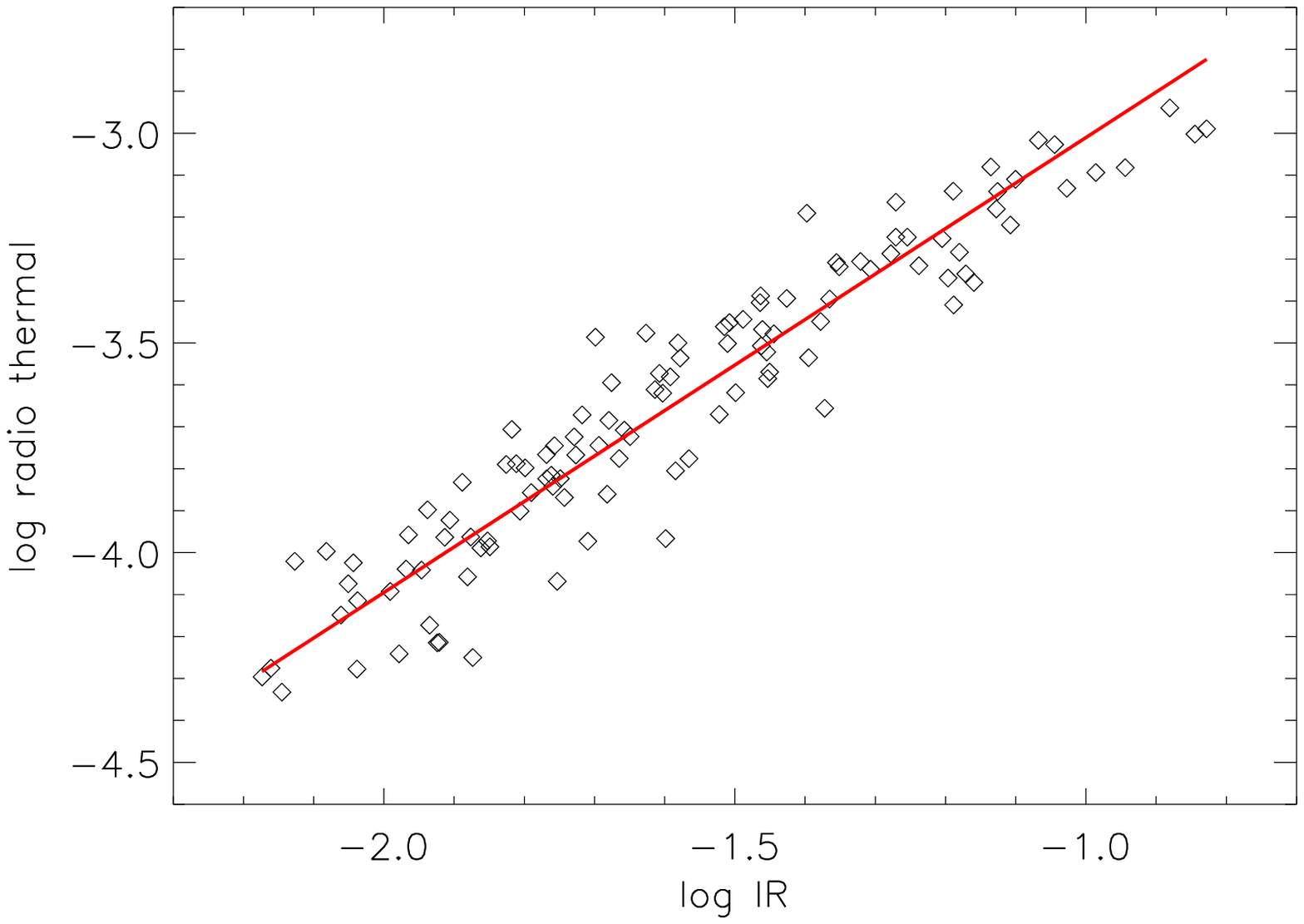}
\includegraphics[width=6.05cm]{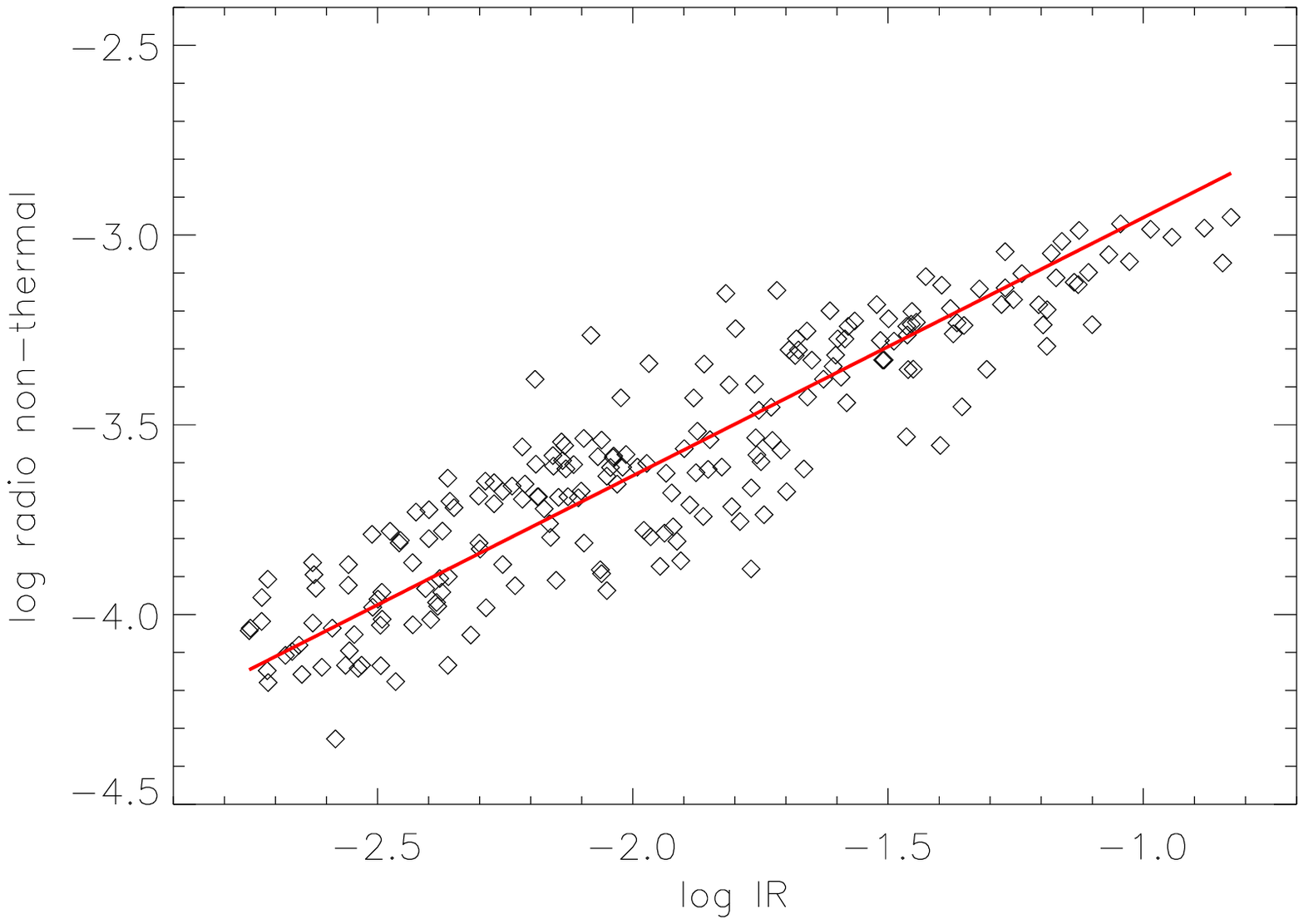}
\includegraphics[width=6.05cm]{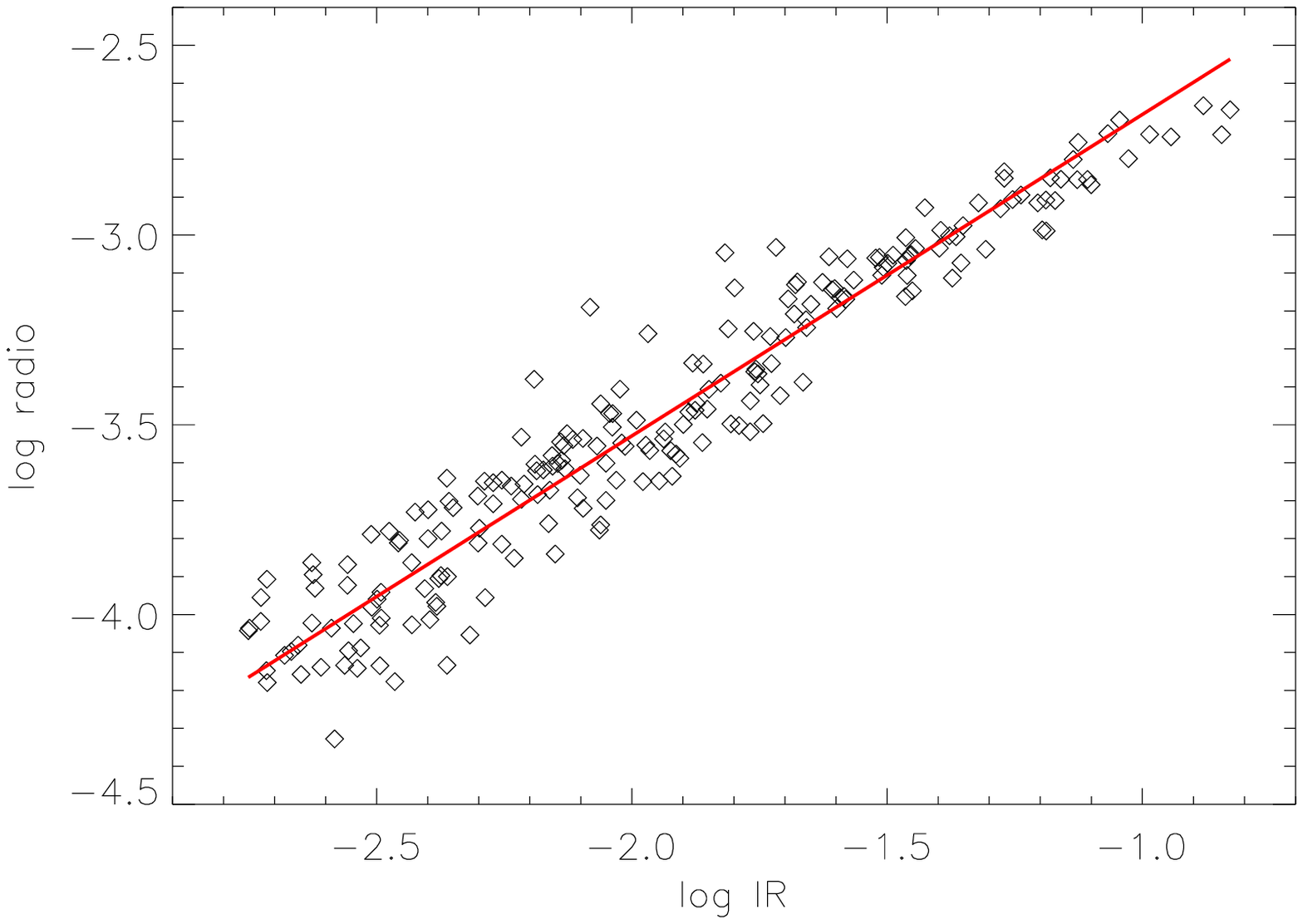}
\caption{
Radio--infrared correlations for radio thermal (left), nonthermal (middle) 
and total (right) emission at 8.46\,GHz at 15\arcsec\ resolution 
within beam--separated regions of NGC\,4254.
} 
\label{f:radfir}

\end{figure*}

The infrared emission and the components of radio intensity at 8.46\,GHz are 
derived in the beam--separated regions within NGC\,4254 at resolution of 15\arcsec\ 
(Fig.~\ref{f:radfir}). The relations are quantified by bisector linear fitting to 
the logarithms of intensities of radio components (Y) and IR fluxes (X). The 
results are given in Table~\ref{t:radfir} together with the Pearson 
correlation coefficients $r$.

The relationship for NGC\,4254 between radio thermal and infrared radiation 
(the slope $\alpha=1.08\pm0.03$) differs considerably from the relation with 
the nonthermal component ($\alpha=0.68\pm0.02$). The slope of 
the total radio/infrared correlation lies in between the above two 
($\alpha=0.85\pm0.02$), closer to the dominant nonthermal component. All the relations are statistically 
highly significant (with correlation coefficients $r\geq 0.90$). The 
relation with the total radio emission is the closest one (Fig.~\ref{f:radfir}). This 
is measured in a quantitative manner by the residual dispersion of the fit. In the case 
of thermal and nonthermal emission these dispersions are 0.16 and 0.14 respectively, 
and just 0.10 for the total emission. 

\begin{table}
\caption[]{The radio--infrared correlation within NGC\,4254: the slopes, correlation 
coefficients $r$, and number of beam--separated regions $N$ used in 
the bisector linear fits to the logarithms of intensities of
different radio components (at 8.46\,GHz) and MIR emission (at $24\,\mu {\rm m}$).
}
\label{t:radfir}
\centering
\begin{tabular}{lllll}
\hline\hline
Y -- X      & slope $\alpha$          & $r$ & $N$  \\
\hline
Radio--thermal -- IR    & $1.08\pm0.03$  & 0.95   & 113\\
Radio--nonthermal -- IR & $0.68\pm0.02$  & 0.90   & 210\\
Radio--total -- IR      & $0.85\pm0.02$  & 0.96   & 210\\ 
\hline 
\end{tabular}
\end{table}

Comparing the obtained slope for the local nonthermal--IR 
correlation in NGC\,4254 with the other galaxies studied to date (which include 
non--cluster galaxies: M31, Hoernes et al. \cite{hoernes98}; M83, Vogler et al. 
\cite{vogler05}; NGC\,6946, Walsh et al. \cite{walsh02}; and the Large Magellanic 
Cloud, Hughes et al. \cite{hughes06}) we notice that 
the slope is always less that one ($\alpha<1$). Hence, this seems to be a general 
phenomenon, not depending on the galactical 
morphology, Hubble type, and cluster membership. However, the slope is $>1$ for the 
{\em global} correlation (Niklas \& Beck \cite{niklas97a}). We propose that
the difference are probably due to propagation effects: the UV photons 
heating the dust can partly escape from the star--forming regions, while the CR
electrons produced in those regions also escape but at different temporal 
and spatial scales. Therefore the nonthermal emission in star--forming regions 
is reduced, but the infrared radiation remains strong. This is well described 
by a phenomenological model in which a radio image appears as a "smeared" version of 
the IR image (Murphy et al. \cite{murphy06}). 

Another explanation for the flat local nonthermal--infrared correlation is the method
of thermal--nonthermal separation. In star--forming regions the nonthermal spectral
index could be flatter than 1.0 due to the direct contribution from young supernova remnants 
(index $\approx 0.5$). This means that the thermal 
emission might be overestimated, and hence the nonthermal emission can be 
underestimated in these regions too. However, this cannot explain why the slope of 
the local correlation between the {\em total} radio and IR emission is also 
flatter than 1.

The comparison of radio thermal--IR relation for NGC\,4254 with the other galaxies 
reveals another trend that the slope of this relation rises systematically 
with IR--wavelengths: from 0.8 and 0.9 for $7\,\mu$m and $15\,\mu$m, 
respectively, for NGC\,6946 (Walsh et al. \cite{walsh02}); through our 1.08 for 
$24\,\mu$m for NGC\,4254; up to 1.17 and 1.31 for $60\,\mu$m and $100\,\mu$m for 
M\,31 (Hoernes et al. \cite{hoernes98}). NGC\,4254 follows this trend, showing 
again that its interaction with the cluster environment does not affect the slope 
of the correlation. At the same time the correlation coefficient reaches its 
maximum at $24\,\mu$m for NGC\,4254 ($r=0.95$), which we recognize as the wavelength 
at which the thermal processes that power radio emission and heat the dust are 
best coupled. In a recent study, Tabatabaei et al. (\cite{taba07a}) make a point 
that beside the dust heating by ionizing UV flux from massive stars 
an aditional dust heating from intermediate-mass and solar mass stars may contribute 
to the emission of the colder dust component seen at longer IR wavelengths.

The observed differences between the correlation slopes of thermal and nonthermal 
emission versus the infrared radiation in NGC\,4254 can lead to a steeper 
{\em total} radio--IR relation for the regions with a higher thermal fraction. 
Such a tendency could explain the recent findings of Hughes et al. (\cite{hughes06}), 
who analyzed the total radio (1.4\,GHz)--IR relation within regions of various 
star--forming intensity of the Large Magellanic Cloud. 

We also compared the {\em global} radio--infrared properties of 
NGC\,4254 with the other nearby field galaxies using the surface brightness at 
4.86\,GHz and $60\,\mu$m: NGC\,4254 fits very well with the general trend of a close 
radio--IR correlation (see Chy\.zy et al. \cite{chyzy06}). Among 19 galaxies of Sc 
type, NGC\,4254 has the second highest surface brightness both in the infrared 
and radio domains. Thus, the observed strong radio emission cannot be 
attributed to the total magnetic field locally compressed by e.g. ram pressure 
of hot ICM. Such a mechanism was suggested to explain the high radio 
luminosity of Coma cluster galaxies (Gavazzi et al. \cite{gavazzi91}). 
We propose that the agent to enhance the SFR in 
NGC\,4254 (most likely the gravitational interaction, Sect.~\ref{s:sfr}) 
increases also (by thermal and nonthermal processes) the infrared 
and radio emission. This explains the strong radio--MIR wavelet correlation 
observed in NGC\,4254 at all spatial scales (Sect.~\ref{s:wave}) as well as 
why the close radio--IR relations revealed by 
non--cluster spirals are valid also for this galaxy. 
Although the external interactions  
do not seem to influence the global radio--IR relation for galaxies 
like NGC\,4254, they do affect the polarized radio intensity. 
Hence, the polarization observations are apparently the key to revealing a 
perturbing agent that is not seen, or barely seen, in other ISM components. 

\subsection{Scenarios of interaction}
\label{s:scenarios}

\subsubsection{Ram pressure}
\label{s:ongoing}

From the available \ion{H}{i} velocity field alone (Fig.~\ref{f:hivelo}) we 
cannot decide whether NGC\,4254 moves inwards or outwards the Virgo 
cluster centre. 
If the galaxy is just entering the Virgo cluster, it can currently experience
ram pressure on the southern disk side by hot ICM. This had been an early idea by 
Cayatte et al. (\cite{cayatte94}) to explain the global N--S asymmetry in the 
distribution of 
the \ion{H}{i} gas (see also Sect.~\ref{s:total}). In this scenario, the southern radio 
polarized ridge could be explained by compression of magnetic field 
at the ISM/ICM interface. Such ram--pressure effects should also spatially 
truncate the H$\alpha$ distribution at the outer disk portion, as shown for a number 
of other Virgo cluster spirals (Koopmann \& Kenney \cite{koopmann04}). In order to check this 
hypothesis we compared H$\alpha$ with the NIR emission 
($3.6\,\mu{\rm m}$) representing the old stellar population in NGC\,4254, 
which should not be affected by ram pressure (Fig.~\ref{f:arms}). Both the distributions 
reveal a complex/disturbed spiral pattern, while there is no 
evidence either for any reduction of H$\alpha$ emission in the southern outer 
disk or for any systematic displacement from the NIR radiation. A similar plot made for e.g. 
the Virgo Cluster spiral NGC\,4522, believed to be ram pressure stripped, 
clearly demonstrates a truncated H$\alpha$ disk as well as extraplanar gas 
(Kenney \& Koopmann\ \cite{kenney99}).

In agreement with that, NGC\,4254 does not reveal any truncation 
in synchrotron emission or in total radio emission even at the longest 
wavelength analysed (Sect.~\ref{s:total}). Our wavelet cross--correlation 
analysis (Sec.~\ref{s:wave}) also indicates similar distributions of 
various ISM gas phases at the largest spatial scales and presence of
an extended \ion{H}{i} envelope. The global distortion of the galactic 
spiral pattern observed in all analysed spectral bands cannot be an argument 
for ram pressure effects as they do not affect stars.

In order to test the importance of ram pressure versus other competing 
forces locally in the southern disk of NGC\,4254 we compared their corresponding 
energy densities. We chose a small region of 15\arcsec\ size in the southern 
polarized ridge, marked by a triangle in Fig.~\ref{f:radio36} (bottom--right panel). 
For calculating the ram pressure $\rho V^2$, we assumed a galaxy velocity 
$V=1000$\,km\,s$^{-1}$ from the hydrodynamical modelling of Vollmer et 
al. (\cite{vollmer05}). The density $\rho$ of the hot cluster's gas was estimated 
from the profiles of Schindler et al. (\cite{schindler}), which for the 
location of NGC\,4254 give $\rho=6.2\times 10^{-5}$\,cm$^{-3}$. 
To calculate the energy density of turbulent gas motions, we assumed a typical 
turbulent velocity of cold gas of about 10\,km\,s$^{-1}$. We took the 
\ion{H}{i} gas density in the ridge and the CO gas density in its eastern part 
(at the edge of the available CO map from Sofue et al. \cite{sofue}) 
as well as 0.2\,kpc disk thickness, which gives an estimate of total gas density 
in the ridge as 3.2\,cm$^{-3}$. 
Finally, from the radio thermal emission we estimated the thermal energy of WIM 
assuming a typical filling factor of gas of about 0.05 and a thermal disk 
thickness of 0.1~kpc and 0.2~kpc.

\begin{table}
\caption{The energy density of different species in the polarized ridge of 
NGC\,4254 in comparison with the ram pressure of ICM and the kinetic energy 
of the infalling \ion{H}{i} clouds.
}
\label{t:energy}
{\centering
\begin{tabular}{ll}
\hline\hline\noalign{\smallskip}
Method   &  Energy density\\
         &
  10$^{-12}$ erg cm$^{-3}$\\
\hline
Ram pressure of ICM                 & 0.5\\
Kinetic energy of \ion{H}{i} plume          & 1.6 -- 3.2\,$^a$ \\ 
Turbul. energy (CO+HI)              & 5.3 \\
Thermal energy of WIM               & 0.8 -- 1.1\,$^b$ \\
\hline
\end{tabular}
\\}
$^a$ -- for various assumptions concerning geometry of the plume\\
$^b$ -- for assumed different thermal disk thickness (0.2 and 0.1\,kpc)\\
\end{table}

The comparison of energy densities presented in Table~\ref{t:energy} shows 
that the cluster ram pressure as well as the thermal energy of WIM are 
considerably (at least five times) lower than the energy in turbulent gas 
motions. Thus we find no evidence for ram pressure operating currently in 
NGC\,4254.

If the galaxy is now receding from the cluster core 
it could pass through it before, experiencing strong ram pressure by the hot 
and dense ICM in the past. If the observed extraplanar \ion{H}{i} 
clouds (Sect~\ref{s:total}) did result from such a past event, we should 
expect a deficiency in \ion{H}{i} gas due to ISM--ICM 
stripping. However, this is not the case (Cayatte et al. \cite{cayatte94}). 
Besides, given the galaxy's current 
large distance from M\,87 ($\approx$1\,Mpc) and typical dispersion speeds of cluster members 
(see Vollmer et al. \cite{vollmer05}) we estimate that this passage could 
have happened about 1.2\,Gyr ago, which is more than two galactic revolutions. 
Such a time is long enough for winding up all the stripped material and   
for the turbulent decay of enhanced magnetic field, which time scale 
is of about $3\times 10^8$\,yr (Widrow \cite{widrow02}). 

\subsubsection{Tidal interaction}
\label{s:tidal}


\begin{figure} 
\centering
\includegraphics[width=5.9cm]{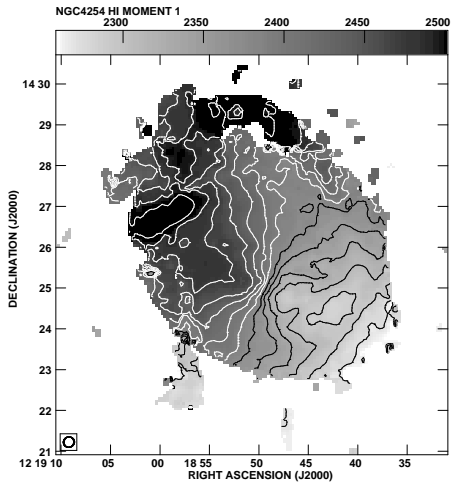}

\caption{
The VLA (archive re--reduced) \ion{H}{i} isovelocity contours of NGC\,4254 (in km\, s$^{-1}$) 
and the greyscale of \ion{H}{i} velocity field. The beam size is 
$15\arcsec$ HPBW. The isovelocity contours correspond to 2.27, 2.29, 2.31, 2.33, 
2.35, 2.37, 2.39, 2.41, 2.43, 2.45, 2.47, 2.49, 2.51 
$\times10^3\,\mathrm{km\,s^{-1}}$ starting from from SW towards NE.
}
\label{f:hivelo}

\end{figure}

If the distortion of spiral structure in NGC\,4254 with its heavy SW 
arm is not of ram pressure origin, it must have been caused by gravitational 
(tidal) interaction. No other process is able to influence the stellar 
orbits to the observed extent. There are two possible candidates for such 
an interaction. The first one is the recently discovered dark galaxy 
VIRGOHI\,21 (Minchin et al. \cite{minchin}). It is located to the north of 
NGC\,4254 at a distance of about 150\,kpc, in the direction of the chain 
of observed weak \ion{H}{i} plumes (see also Fig.~\ref{f:hiint} in the 
Online Material). However, 
its large mass (10$^{11}\mathrm{M}_{\sun}$) speculated from its rotation 
curve is yet to be confirmed. The second candidate is the spiral galaxy 
NGC\,4262 with a well--established large mass. It is also located north of 
NGC\,4254, but it is not disturbed and the chain of \ion{H}{i} blobs do 
not seem to be related to it (Minchin et al. \cite{minchin}). According to 
numerical simulations (Vollmer et al. \cite{vollmer05}) NGC\,4254 might 
have suffered from a gravitational encounter some $10^8$\,yr ago. 

Another argument for gravitational interactions is provided by our SFR 
analysis (Sect.~\ref{s:sfr}). The galaxy's mean SFR is enhanced but not 
spatially truncated. This kind of external influence on 
the Virgo cluster spirals is indeed well recognized as resulting from tidal 
interactions (Koopman \& Kenney \cite{koopmann04}). A SFR enhancement, but 
to a larger extent, was also observed in the gravitationally interacting 
Antennae galaxies (Chy\.zy \& Beck \cite{chyzy04}). The steep global 
radio spectral index in NGC\,4254 of 0.8 (Sect.~\ref{s:spectral}) indicates 
an evolved population of star--forming regions, in agreement with a star 
formation triggered by gravitational encounter occurring a considerable 
time ago. 

The global N--S asymmetry in the \ion{H}{i} gas seen in the low--resolution 
(40\arcsec) data by Cayatte et al. (\cite{cayatte94}) appears at higher 
resolution of 15\arcsec\ to be composed of disk and out--of--disk components 
(Fig.~\ref{f:hivelo}). The external gas can be well discerned in the NE as a huge 
\ion{H}{i} arc of velocity different from that of gas in the disk (recognized 
also by Phookun et al. \cite{phookun93}). Our second moment map of the 
\ion{H}{i} velocity field (Fig.~\ref{f:disp}, available in the Online Material) 
does show an enhanced velocity dispersion to above 20\,km\,s$^{-1}$ at the 
point where the arc enters the disk from the north (RA=$12^{\mathrm{h}} 
18^{\mathrm{m}}	44^{\mathrm{s}}$, Dec=$14\degr 28\arcmin 0\arcsec$). This 
region is rather small ($40\arcsec\times40\arcsec$) and resembles the place in 
the southern disk where a single elongated \ion{H}{i} blob starts to penetrate 
the disk. This favours the idea of the arc as a debris feature of a past 
gravitational encounter. In the case of gas pushed out from the disk by ram 
pressure we should not expect a sudden but a gentle and wide distortion of 
the velocity pattern, as e.g. can be seen in the \ion{H}{i} tail of NGC\,4654 
(Phookun and Mundy \cite{phookun95}).

Support comes from the inspection of recent Fabry--Perot observations in the 
H$\alpha$ line (Chemin et al. \cite{chemin06}) showing a very regular 
disk rotation without any counterpart of the \ion{H}{i} arc. 
Similarly, we did not find any associated magnetized plasma traced by the extended 
radio envelope at 1.43\,GHz (Sect.~\ref{s:total}) or any other ISM component.  
Thus, the \ion{H}{i} arc and the global N--S asymmetry of \ion{H}{i} gas is better 
explained by gravitational interaction than by the  
ram pressure hypothesis.

\subsubsection{The puzzle of the polarized emission}
\label{s:puzzle}

NGC\,4254 seems to belong to the class of  
`young' Virgo cluster members, which recently experienced 
a gravitational encounter at the cluster's periphery, which perturbed its spiral 
arms by tidal forces, thus triggering a burst of star--formation that is 
still at a high level (Sect.~\ref{s:sfr}) and maintains strong radio and 
infrared emissions (Sect.~\ref{s:fir}). 
The southern polarized ridge is, however, much more difficult to 
account for as it has no visible counterpart in the other spectral ranges 
(Sect.~\ref{s:polar}), and partly because such a strong magnetic structure 
lying {\em downstream} the spiral density wave has been never observed in any 
other galaxy. 

Within the framework of gravitational interaction, the enhanced polarized 
emission in the southern disk could potentially be caused by compressing 
plasma and magnetic field by the tidally stripped \ion{H}{i} gas falling now back
onto the disk. In order to check this possibility we calculated the 
ram pressure exerted by the large southern blob visible in 
Fig.~\ref{f:hiint}. The blob is probably entering the disk with 
relative velocity of about 100\,km\,s$^{-1}$ (Fig.~\ref{f:hivelo}). 
The mean gas density in the blob is 0.08\,cm$^{-3}$ and 0.04\,cm$^{-3}$ 
for the gas thickness values of 1\,kpc and 2\,kpc, respectively. According 
to our estimates (Table~\ref{t:energy}), the energy associated with  
returning cold gas is comparable to the thermal energy but two times 
lower than the turbulent energy in the disk. The blob does not seem to 
affect either the strength or the orientation of magnetic field in the disk
(Fig.~\ref{f:radio36}), which may indicate that it moves along the orbit inclined 
to the disk. Both the arguments seem to exclude the hypothesis of plasma compression 
by returning cold gas, but the definite answer could bring only a careful MHD 
modelling of this interaction. 

We consider also another possibility that the same tidal forces that were 
to disturb the galactic stellar and gaseous content could 
stretch and shear magnetized plasma along the SW spiral arm as well. This 
should result in transforming the random (isotropic) magnetic field 
component to an anisotropic one and enhance locally the polarized radio 
emission (cf. Sokoloff et al. \cite{sokoloff98}). 
As the polarized emission constitutes just about 24\% 
of the total one, it is obvious why such effects cannot be found 
in total radio emission and other ISM components. 
Whether the ridge can be actually caused by shearing cannot be established without a 
quantitative modelling of magnetic field components. This 
will be attempted in Paper II.

We still cannot exclude the possibility that some {\em weak} ram pressure 
effects from hot ICM gas are still sufficient to modify/compress 
magnetic field in the southern ridge, while not affecting other ISM components. 
This would mean that magnetic fields in NGC\,4254 are much more sensitive 
to compressional forces than all other ISM phases. This hypothesis can be 
tested by a thorough MHD modelling of plasma, which is quite challenging as it requires
an accurate reconstruction of the gravitational encounter, the pattern of 
spiral density waves and their influence on the different components of 
magnetic field. 

In contrast to the radio thermal component, which highly 
correlates with the other species of thermal origin (Sect.~\ref{s:wave}),
the radio polarized signal in NGC\,4254 is not closely connected with the other 
ISM agents and in fact even anticorrelates with most of them in the
southern polarized ridge. 
Follow--up studies of other galaxies in the Virgo cluster are necessary to 
fully understand the role of magnetic field and environmental 
effects in cluster galaxies. The case of NGC\,4254 shows that 
the polarized signal provides additional information on MHD 
processes acting on magnetized plasma during the galaxy's evolution, which cannot 
be obtained from any other ISM component. The presented analysis also shows that 
only high resolution (1\,kpc--scale) and multifrequency studies are appropriate 
for discerning various environmental processes influencing galaxies 
within clusters. 

\section{Summary}
\label{summary}

We present the first comprehensive investigation of the weakly perturbed 
Virgo cluster spiral NGC\,4254. We performed VLA radio polarimetric 
observations at 8.46\,GHz, 4.86\,GHz and 1.43\,GHz, with sensitivity
to extended structures enhanced by single dish 100--m Effelsberg observations 
at the former two frequencies. We also carried out observations of hot gas 
components in X--ray and UV emission with the XMM--Newton satellite and 
re--analyzed \ion{H}{i} observations obtained by Phookun 
et al. (\cite{phookun93}). On the basis of our data and images in other spectral 
domains, we investigate the interrelations between different gas 
phases as well as the influence of the cluster 
environment on the galaxy's morphology, especially in the radio domain.  

We found the following:

\begin{enumerate}

\item
The distributions of total radio intensity at 4.86\,GHz and 8.46\,GHz are 
disturbed. In the north, the emission is more diffuse and extends two times 
further than in the south. It follows the galaxy's optical images, 
and thus the stellar and gaseous content of the unusual three--arm spiral 
structure. The strongest emission comes from the 
central disk region, where a bar-like structure can be discerned at the highest 
($7\farcs 5$) radio resolution available, corresponding to a similar feature 
appearing in CO emission (Sofue et al. \cite{sofue}). 

\item The polarized part of synchrotron emission at 8.46\,GHz and 4.86\,GHz
is highly asymmetric and shows a strange, sharp and bright, ridge in the 
southern disk {\em shifted outwards} of the optical SW spiral arm, downstream 
the density wave. There is a high degree of polarization in this region, reaching 
locally 40\% (and mean value of 24\%). Three other polarized ridges are 
interlaced with adjacent optical spiral arms, while still another one is aligned with them. 
This complex mixture of magnetic field patterns as well as the strong global asymmetry 
of polarized emission is very unusual among spiral galaxies observed so far at 
such high radio frequencies. 

\item The orientations of magnetic field vectors vary strongly over the 
galaxy, from almost zero apparent magnetic pitch angles in the south 
up to 30\degr-40\degr\ in the northern part, but they are mainly aligned 
(within 20\degr) with orientations of optical arms. We report the discovery 
of a large magnetized radio envelope of NGC\,4254 at 1.43\,GHz 
with ordered magnetic fields extending further outwards (4\,kpc) than the 
warm and hot gas.

\item Our XMM-Newton data reveal some soft X-ray emission tightly associated with 
star-forming regions almost from the whole galactic disk. The 
X-ray distribution is disturbed along a similar pattern as radio and optical emission and 
shows no out-of-disk emission or shocks that could imply strong ram 
pressure forces.

\item
The wavelet decomposition of images reveals the highest wavelet 
cross-correlation between various species dominated by thermal 
processes, especially between radio thermal and $24\,\mu$m mid-infrared 
emission (correlation coefficients $r_w\ge 0.88$ at all inspected spatial scales). 
However, the polarized emission anticorrelates ($r_w=-0.4$) with the radio 
thermal emission as well as the X-ray emission, manifesting the displacement of magnetic 
arms from the optical ones. These results are quite similar to the 
non-cluster spiral NGC\,6946 (Frick et al. \cite{frick01}).

\item 
For the first time we use the radio thermal emission at 8.46\,GHz to derive an 
extinction-free SFR distribution across the galaxy. We find higher 
extinction values in more H$\alpha$ luminous star-forming regions with a power-law 
slope of 0.83. The estimated mean SFR over the galactic disk of $0.026\,\mathrm{M}_{\sun}\,
\mathrm{yr}^{-1}\,\mathrm{kpc}^{-2}$ (corresponding to a global SFR of 
$8.4\,\mathrm{M}_{\sun}~\mathrm{yr}^{-1}$) is higher than in typical 
field spirals of the same Hubble (Sc) type. Contrary to the main trend among Virgo Cluster 
spirals, the star-forming disk in NGC\,4254 is neither spatially truncated, nor shifted 
from the old stellar population, which excludes strong ram-pressure stripping 
by hot ICM.

\item
Investigating for the first time the {\em local} 
radio-IR relation for a disturbed cluster spiral, we found for NGC\,4254 a strong correlation ($r\ge0.90$) 
of radio thermal, nonthermal, and total emission at 8.46\,GHz with $24\,\mu$m 
(MIR) radiation. The obtained flat slope ($<1$) for the local 
{\em nonthermal}--IR relation differs from the {\em thermal}--IR one ($\ge1$). We argue 
that the thermal process that powers the radio emission is best coupled with  
dust heating at $24\,\mu$m. We find that NGC\,4254 is consistent with the {\em global} 
radio-IR relationship determined for the field galaxies.

\item 
The idea of ram pressure of hot ICM acting on the galaxy disk is questionable 
in view of the analyzed multifrequency data. We argue that the spiral arm pattern 
has been perturbed by tidal forces, which both induced a strong star formation rate and 
sheared the magnetic field in the southern disk along the optically bright SW spiral arm, leading 
to the strong polarized ridge. In order to make sure if such process could be common among 
other disturbed spirals in the Virgo Cluster, similar comprehensive and 
multifrequency studies of them are desirable.

\end{enumerate}

\begin{acknowledgements}
The Authors wish to express their thanks to Dr Igor Patrikeev for 
help in the wavelet calculations and to Dr J. Knapen for assistance in 
calibration of the H$\alpha$ image. We are also grateful to Prof. M. Urbanik, Dr 
M. Krause and the anonymous referee for valuable comments. This work was supported by a grant 
from the Polish Research Committee (KBN), grant no. PB249/P03/2001/21.
\end{acknowledgements}

\onlfig{14}{

\begin{figure} 
\centering
\includegraphics[width=8cm]{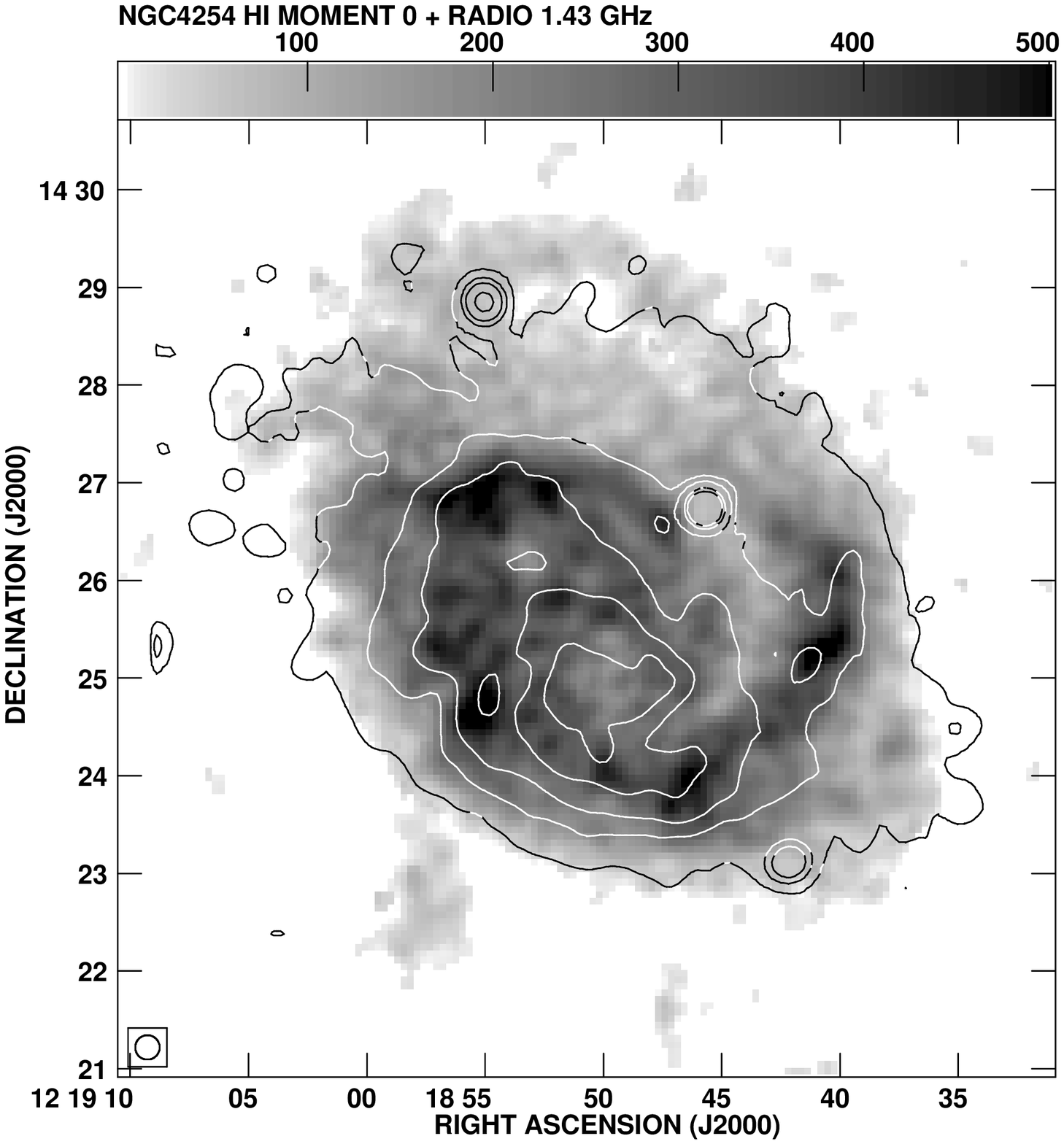}

\caption{
VLA (archive re-reduced) \ion{H}{i} intensity of NGC\,4254 in greyscale (in Jy 
b.a.$^{-1}$\,km\,s$^{-1}$) and the radio continuum emission at 1.43\,GHz in contours. 
The beam size is $15\arcsec$ HPBW.
}
\label{f:hiint}
\end{figure}
}


\onlfig{15}{
\begin{figure}
\includegraphics[width=8cm]{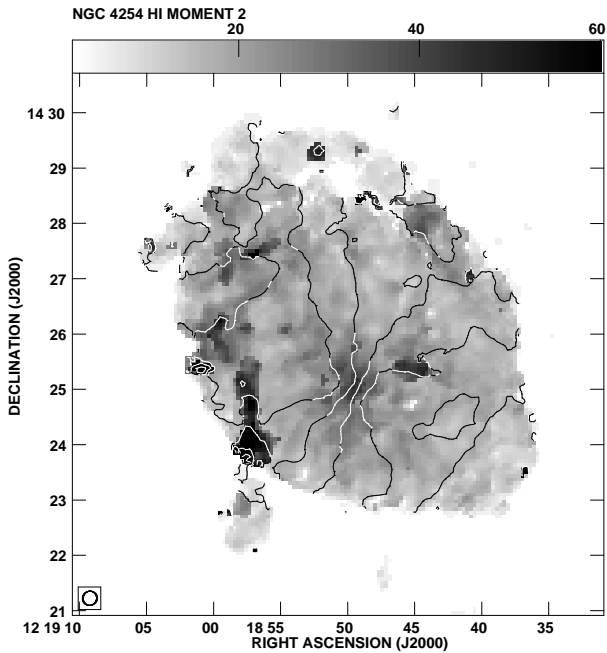}
\caption{
Velocity dispersion of \ion{H}{i} gas in NGC\,4254 (in km\,s$^{-1}$) in 
greyscale with isovelocity contours at 2.29, 2.33, 2.37, 2.41 $\times10^3\,
\mathrm{km\, s^{-1}}$ (see also Fig.~\ref{f:hivelo}). The VLA (archive 
re-reduced) data have the beam size of $15\arcsec$ HPBW.
}
\label{f:disp}
\end{figure}
}

\end{document}